\documentclass[traditabstract, twocolumns]{aa}
\usepackage{graphicx}
\usepackage{xcolor}
\usepackage{txfonts}
\usepackage{makecell}
\usepackage{hyperref}
\hypersetup{
    colorlinks=true,
    linkcolor=blue,
    filecolor=magenta,      
    urlcolor=blue,
    citecolor=blue,
    pdfpagemode=FullScreen,
    }
\usepackage{graphicx}	% Including figure files
\usepackage{amsmath}	% Advanced maths commands
\usepackage{caption}
\usepackage{subcaption}
\usepackage{multirow}
\usepackage{pifont}% http://ctan.org/pkg/pifont
\usepackage[normalem]{ulem}
\usepackage{booktabs}
\newcommand{\kmsmpc}{km\:s^{-1}\:Mpc^{-1}}

\defcitealias{Bruzual2003}{BC16}
\defcitealias{adamo_bound_2024}{A24}
\defcitealias{messa_jwst_2026}{M26}

\begin{document}

   \title{Cosmic CORALS: Timing the Universe with high-z star clusters}
   
   \author{Elena Tomasetti
          \inst{1,2}\thanks{e-mail: elena.tomasetti2@unibo.it}
          \and 
          Martin Millon\inst{3,4}
          \and
          Licia Verde\inst{5,6}
          \and
          Frédéric Courbin\inst{5,6,7}
          \and
          Raul Jimenez\inst{5,6}
          \and
          Michele Moresco\inst{1,2}
          \and
          Carmela Lardo\inst{1,2}
          \and
          Andrea Cimatti\inst{1,2}}

   \institute{Dipartimento di Fisica e Astronomia ``Augusto Righi”–Universit\`a di Bologna, via Piero Gobetti 93/2, I-40129 Bologna, Italy.
         \and
             INAF - Osservatorio di Astrofisica e Scienza dello Spazio di Bologna, via Piero Gobetti 93/3, I-40129 Bologna, Italy.           
        \and
        D\'epartement de Physique Th\'eorique, Universit\'e de Gen\`eve, 24 quai Ernest-Ansermet, CH-1211 Gen\`eve 4, Switzerland 
        \and 
        Institute for Particle Physics and Astrophysics, ETH Zurich,
        Wolfgang-Pauli-Strasse 27, CH-8093 Zurich, Switzerland
        \and
            ICC, University of Barcelona, Marti i Franques 1, 08028 Barcelona, Spain.
        \and
            ICREA, Pg. Lluis Companys 23, Barcelona, 08010, Spain. 
             \and
            IEEC, Institut d'Estudis Espacials de Catalunya, Edifici RDIT, Campus UPC, Castelldefels, 08860 Barcelona, Spain
        }

    \titlerunning{Cosmic CORALS}
    \authorrunning{E. Tomasetti et al.}

\abstract{In this work, we explore the potential of anchoring the age-redshift relation across cosmic time by probing the oldest star clusters at high redshift, now observed thanks to the James Webb Space Telescope in strongly lensed fields. As a case study, we consider one of the highest-redshift systems observed, the Cosmic Gems arc at $z=9.625$. We perform image deconvolution of multi-band JWST imaging, identifying a total of 20 point sources along the arc. We derive the stellar ages through a cosmology-independent spectral energy distribution (SED) fitting framework, ensuring that these measurements can be used as unbiased cosmological anchors. By combining these high-z systems with state-of-the-art local globular cluster ages, we perform a joint Bayesian fit to the age-redshift relation in a flat $\Lambda$CDM model, measuring $H_0=70^{+27}_{-16}\ \rm{km\ s^{-1}\ Mpc^{-1}}$ and $\Omega_m=0.33^{+0.37}_{-0.21}$. While these constraints are still loose, we show that the slope of the degeneracy, a power-law in the $\Omega_m - H_0$ plane, is highly dependent on the redshift of the sources, becoming shallower as redshift increases. 
Leveraging this geometric rotation, we present forecasts showing that a future sample of $\sim 300$ lensed proto-globular clusters well-distributed up to $z \approx 10$ could tighten the statistical precision to $4\%$ on $H_0$ and $11\%$ on $\Omega_m$, competitive with and independent of methods currently in use. The present work, therefore, represents a new avenue in cosmology and comes at a timely moment, when JWST observes high-redshift lensed star clusters routinely, Euclid and the Nancy Grace Roman Space Telescope uncover new strong lensing fields, and close to the start of operation of the ESO Extremely Large Telescope.

}


   \maketitle
%
%-------------------------------------------------------------------

\section{Introduction}
\label{sec:1Intro}

Since the launch of the James Webb Space Telescope (JWST), a new window has opened on the distant Universe, unveiling the earliest epochs of galaxy formation. The combination of this unprecedented facility with the magnifying power of gravitational lenses, has further enhanced the possibility of observing objects that would be too faint to see at cosmological distances, like star clusters \citep{vanzella_early_2022, vanzella_jwstnircam_2023,Mowla2022,Mowla2024,adamo_bound_2024}, or even single stars \citep{kelly_extreme_2018,welch_highly_2022}. This not only allows for a direct investigation of how the local-Universe structures have formed and evolved, but also provides a unique opportunity to interpret these findings from a cosmological perspective.

Globular clusters (GCs) have an important role in cosmology, as they represent the oldest structures we can observe locally, and for which ages can be measured robustly, representing the most direct constraint on the minimum age of the Universe \citep{boylan-kolchin_uncertain_2021,Cimatti2023,tomasetti_globular_2025,valcin_age_2025,valcin_age_2026}. The age of the Universe, $t_{\rm{U}}$, is linked to other key cosmological parameters, being highly sensitive to the Hubble constant, $H_0$. Specifically, within a flat $\Lambda$CDM framework, the current $H_0$ tension \citep[see, e.g.,][]{di_valentino_cosmoverse_2025} introduces a discrepancy of over 1 Gyr in the absolute age of the Universe at $z=0$ for a fixed matter density parameter, $\Omega_m=0.3$. Consequently, precise age measurements can provide an independent cross-check for the current $H_0$ debate.

With JWST's ability to resolve distant stellar clusters, we are now entering an era in which we can potentially detect the progenitors of local GCs, the oldest structures in the young Universe. The direct connection between these high-redshift compact structures and present-day GCs is now being actively investigated and mapped through both observations and state-of-the-art cosmological simulations \citep[see, e.g.,][]{pfeffer_e-mosaics_2018,pozzetti_search_2019,pfeffer_comparing_2025,giunchi_dynamical_2025,claeyssens_first_2026,della_croce_evolution_2026}.
By anchoring the age of the Universe at multiple epochs, these high-redshift stellar systems serve as unique probes to test the consistency of our cosmological model across cosmic time. 

In \citet{tomasetti_time_2025}, we proposed this idea for the first time, investigating the lensed globular cluster candidates of the Sparkler galaxy \citep[z=1.38,][]{Mowla2022} and deriving their ages in a cosmology-independent way. Here, we want to take it a step further. 
We extend the sample to one of the highest redshift known lensed stellar systems, the Cosmic Gems \citep[z=9.625][]{adamo_bound_2024,messa_jwst_2026, vanzella_z_2026}. Together with the local GCs ages estimates and those of the Sparkler's GCs, they provide the first reconstruction of $t_{\rm{U}}$ up to the very first phases of the Universe's evolution. 

This effort marks the launch of \textsc{CORALS} (Clusters of stars as Observational Records of cosmic Aging at Lookback timeS), a project dedicated to exploring the use of high-redshift star clusters as cosmological probes. The few sources considered in this work, indeed, as remarkable as they are, are probably just the tip of the iceberg: the number of observed strongly lensed globular clusters is expected to increase dramatically in the near future thanks to the synergistic effort of wide-field space surveys like Euclid and the Nancy Grace Roman Space Telescope, combined with the deep targeted imaging of JWST and, looking ahead, the unprecedented resolution of the Extremely Large Telescope (ELT). In this work, we not only exploit the Gems clusters for cosmological applications, but we also explore the future of the \textsc{CORALS} project by extending this approach to a statistically significant sample of proto-GCs, spanning a wide redshift range.

In Sect.~\ref{sec:2DATA} we describe the JWST imaging data used in this work and how we extract reliable multi-band photometry of the Gems. Section~\ref{sec:3METHOD} presents how this photometry is used to infer their ages independently of cosmology. These ages are then used in Sect.~\ref{sec:cosmo} to infer cosmological constraints, namely on $H_0$ and $\Omega_m$ in a flat $\Lambda$CDM universe. Section~\ref{sec:4cosmo_fit_mocks} extends our pilot study with simulations to forecast future constraints with larger samples and to illustrate the sensitivity of the results with redshift. Our conclusions are exposed in Sect.~\ref{sec:5CONCLUSIONS}.

%##############se#######  DATA #####################

\section{Data and photometry with \texttt{STARRED}}
\label{sec:2DATA}
JWST imaging for the Cosmic Gems Arc is available in eight NIRCam bands: F090W, F115W, F150W, and F200W in the Short Wavelength (SW) channel, and F277W, F356W, F410M, and F444W in the Long Wavelength (LW) channel \citep[GO 4212, P.I. L. Bradley, see][]{bradley_unveiling_2025}. \citet{adamo_bound_2024} \citepalias[][hereafter]{adamo_bound_2024} first identified five distinct bright star clusters, designated sources A through E, as massive ($\sim10^6\ M_\odot$), compact ($\sim1\text{ pc}$) star clusters. This system was subsequently followed up with JWST/NIRSpec IFU spectroscopy by \citet{messa_jwst_2026} \citepalias[][hereafter]{messa_jwst_2026}, who refined the spectroscopic redshift of the arc, $z = 9.625 \pm 0.002$, and identified it as a post-starburst system.

Because the background galaxy intersects the lensing caustic, each of these sources is symmetrically split into two mirrored counter-images, yielding a total of ten individual point-like sources. Due to their close proximity to the lensing caustic, these sources experience extreme gravitational lensing, with local magnification factors ranging from tens to several hundreds \citepalias{messa_jwst_2026}. In addition to these primary clusters, the arc shows two extended tails characterized by at least four fainter point sources, compatible with star clusters, labelled F-I in \citepalias[][]{messa_jwst_2026} in superposition to diffuse emission. 

\begin{figure}[t]
    \centering
    \includegraphics[width=0.92\linewidth]{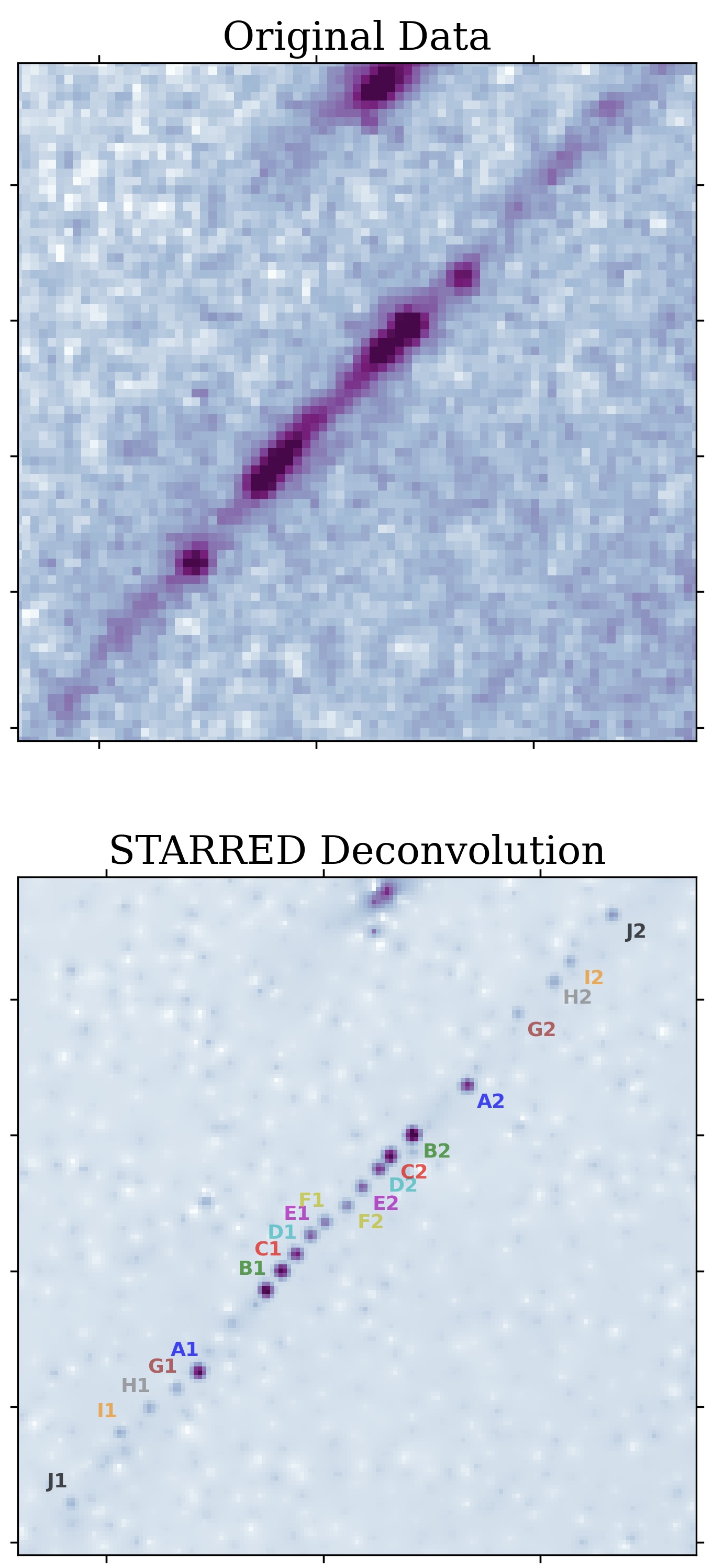}
    \caption{\textit{Top}: F200W NIRCam observation of the Cosmic Gems arc from \citet{bradley_unveiling_2025}. The spatial resolution is 0.066\arcsec\, and the pixel size is 0.031\arcsec. \textit{Bottom:} \texttt{STARRED} deconvolution for the same filter, along with the labels for each of the 20 point sources. The new pixel size is 0.016\arcsec\, and the resolution is 0.031\arcsec\ Full-Width-Half-Maximum.}
    \label{fig:deconvolution}
\end{figure}

Given the presence of blended clusters within the arc, and to effectively isolate the oldest components of this system, we perform deconvolution photometry with \texttt{STARRED} \citep{millon_image_2024, Michalewicz2023}. This framework utilizes a two-channel pipeline that efficiently separates the emission of compact point sources from the underlying extended background.

Fig.~\ref{fig:deconvolution} compares the original JWST image and the deconvolved one in the F200W filter: the labels A-E indicate the clusters reported in the literature.
While a comprehensive description of the deconvolution process is deferred to Appendix~\ref{sec:app_deconv}, here we highlight that this algorithm allows us to uncover a sixth pair of mirrored point sources in the main body of the arc, which we designate as source `F', as an extension to the published list. These innermost sources could correspond to the compact line-emitting region detected by \citetalias{messa_jwst_2026}, located near the critical line and previously undetected in broad-band imaging. In addition, \texttt{STARRED} successfully extracts the photometric emission of the four additional faint sources located within the diffuse tails, which we identify as sources G to J. This yields a total of 20 point sources, all shown in Fig.~\ref{fig:deconvolution}.
%In Fig. \ref{fig:deconvolution} we report both the original image and the deconvolved one in the F200W filter, with the respective identifiers besides all the 20 point sources. 

%%%%%%%%%%%%%%%  METHOD #########################

\begin{figure*}[t] 
     \centering
     \includegraphics[width=0.98\textwidth]{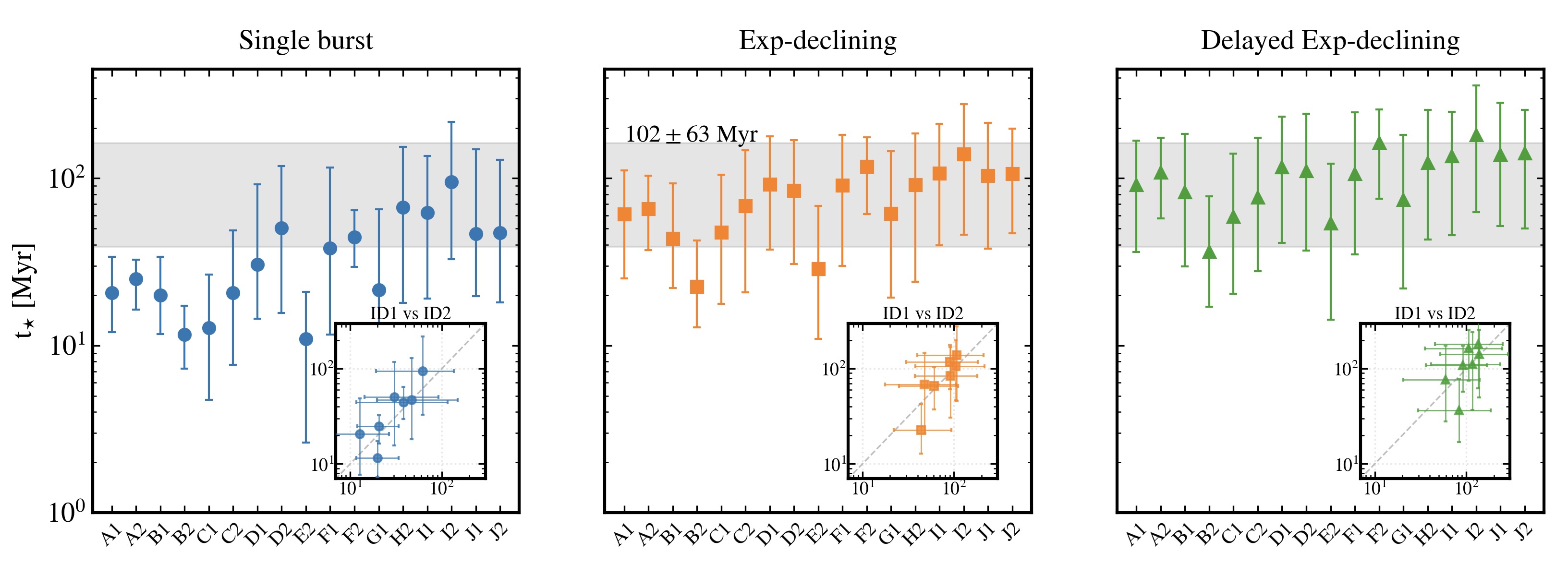}
         \caption{Visual summary of the recovered ages in the three SFH scenarios for the 10 objects. In each panel, the inset shows the 1-to-1 age relation of counter images (e.g., A1 vs A2) when both are available. Instantaneous (single burst) SFH returns younger ages with smaller errors. Exponentially declining and delayed exponential return increasingly older ages, but consistent within the errors. We adopt the exponential declining SFH as our fiducial age determination. The shaded grey area represents the age range covered by the oldest half of the sample in this configuration, $t_{\star}=102 \pm 63$ Myr, and is reported in all panels.} \label{fig:agesall}
\end{figure*}

All clusters are consistently and independently detected in the F150W and F200W bands, exhibiting signal-to-noise ratios (SNR) ranging from 4 to 45. In contrast, the lower spatial resolution of the LW channel limits proper identification in these bands; thus, the positions of the sources identified in the SW filters are used as priors to extract the photometry in the LW bands. To ensure SED reliability, we apply a quality cut, discarding sources with three or more unreliable detections, namely E1, H1, and G2. Nevertheless, because at least one counter-image of each lensed pair passes our quality cut, SED can be successfully extracted for all 10 individual point-sources.

\section{Age and metallicity of the Gems proto-GCs}
\label{sec:3METHOD}

We derive the age, overall metallicity, $Z/Z_\odot$, and dust attenuation, $A_V$, for all 10 detected sources using the public SED-fitting code \texttt{BAGPIPES} \citep{Carnall2018}. The Stellar Population Synthesis (SPS) models used correspond to \citet{Bruzual2003}, updated in 2016, which adopt a \citet{kroupa_variation_2001} Initial Mass Function (IMF). We explore three different Star Formation Histories (SFH), namely: single burst, exponentially declining, and delayed exponentially declining. Given the short timescales measured for these objects in the local Universe \citep[see, e.g.,][]{Bastian2018,martocchia_search_2018},
we limit the duration of star formation episodes to 100 Myr.
Further details and extended results of the fitting process are presented in Appendix~\ref{sec:app_FSF}.

Since our primary goal is to exploit the measured stellar ages as cosmological proxies, we need to ensure that our constraints are entirely independent of any assumed cosmological model. Consequently, following the methodology established in our previous studies \citep{tomasetti_time_2025, tomasetti_globular_2025}, we explicitly remove the cosmological prior for the age during the fitting process. Importantly, for each object, we derive two age estimates that have different meanings and use:

\begin{itemize}

\item {The standard stellar age, $t_{\star}$}, which is the time elapsed since the very onset of star formation within the chosen parametric model. It is the preferred metric in the cosmological context as it directly anchors the system to the cosmic epoch when the star-formation episodes began, hence providing a direct link to its formation redshift.
\vskip 3pt
\item {The mass-weighted age, $t_{\rm mw}$}, which reflects the average age of the bulk of the stellar mass, identifying the peak of the SFH. It is by definition smaller than or equal to the standard stellar age.

\end{itemize}

In this work, we adopt $t_{\star}$ for the cosmological analysis, while $t_{\rm mw}$ is used instead for comparisons with results taken from the literature, to ensure consistency.

\subsection{How old are the Gems?}\label{sec:Gems_ages}

The derived values of the standard age are sensitive to the assumed SFH, showing systematic, yet limited, variations across the three scenarios. 
This is visualized in Fig.~\ref{fig:agesall}, and we refer the reader to Table~\ref{tab:sed_results} in Appendix~\ref{sec:app_FSF} for the full results.

Assuming an instantaneous star-formation event, the single-burst model implies that all stars form into a single initial episode. This yields the youngest age estimates (and smaller statistical uncertainties) compared with other assumptions. The central sources (A–F) are characterized by ages from just 11 Myr up to 50 Myr, while the tail sources (G–J) emerge as the oldest, peaking at 95 Myr. In the non-instantaneous exponential and delayed models, the delay $\tau$ converges to a remarkably consistent window of $\sim$60–70 Myr for all sources.
This pushes the absolute onset of star formation systematically further back in time. In turn, absolute ages also increase, but we consistently find the same age gradient between the central sources and the tail sources, as shown in Fig. \ref{fig:agesall}. 
This spatial layout is accompanied by a consistently low metallicity across all scenarios, with median values concentrated between 0.2\% $Z_\odot$ and 1.6\% $Z_\odot$.

Because of the large statistical uncertainties associated with these measurements, these values should be interpreted as indicative ranges rather than precise measurements: while the data statistically favour a young, low-metallicity system, the constraining power of just 6 informative photometric points does not allow the SED fitting to rule out scenarios involving much earlier star-formation onsets or higher chemical enrichment. 

As just discussed, and illustrated in Fig.~\ref{fig:agesall}, the inferred ages depend on the assumed SFH, increasing in value and uncertainty with the complexity of the functional form.
While the true formation history of these objects cannot be uniquely constrained from integrated photometry, the exponentially declining model represents a robust middle ground. The single-burst imposes the rigid, unphysical limit of instantaneous formation, which minimises degeneracies but mostly tracks the peak of the star formation, rather than its onset. Conversely, the delayed-exponential model enforces a gradual rising phase, which pushes the onset to older lookback times. The exponentially declining SFH allows star formation to extend over a finite timescale without imposing a slow-rise phase, which may be unfit for very rapid star-forming episodes. It therefore represents a physically plausible reference choice. 

Crucially, as shown in Figure~\ref{fig:agesall}, the age estimates from all three models remain mutually consistent within $1\sigma$. Such an agreement demonstrates that the age of the system is robustly pinned down within a tight window of approximately 100 Myr, regardless of the specific SFH prescription. Achieving this level of precision at such high redshift and independently of cosmology represents a powerful result, demonstrating that even with the degeneracies inherent to photometric fitting, we can robustly place tight boundaries on the timing of star formation for these early structures. This consistency also holds when benchmarking our results against different literature frameworks. Specifically, our derived ages show remarkable agreement with the independent estimates from \citetalias{messa_jwst_2026}, with an average difference of only $\sim 20$ Myr (see Appendix~\ref{sec:App_comparison} for a complete comparison).

To derive a single, robust time constraint for each stellar system, we compute the average of the standard ages, $t_\star$, resulting from the SED fitting of the two counter-images, weighted on the associated errors. 
For the cosmological analysis, using the so-computed ages, we select the oldest half of the sample, since these systems provide the strongest lower limits on the time available for star formation before the observed epoch. The division is made at the median age ($\langle t_\star \rangle = 76$ Myr), rather than through an arbitrary absolute-age threshold, and therefore retains a statistically meaningful subsample while isolating the objects with the greatest cosmological leverage (namely sources D, F, H, I, J). The combined age distribution of this oldest half of the sample is $t_\star = 102 \pm 63$ Myr, and is shown as a grey shaded horizontal band in Fig. \ref{fig:agesall}.

\section{Cosmological implications}\label{sec:cosmo}

\begin{figure*}[h!] 
     \centering
     %\hfill 
     \begin{subfigure}[b]{0.5\textwidth}
         \centering
         \includegraphics[width=\textwidth]{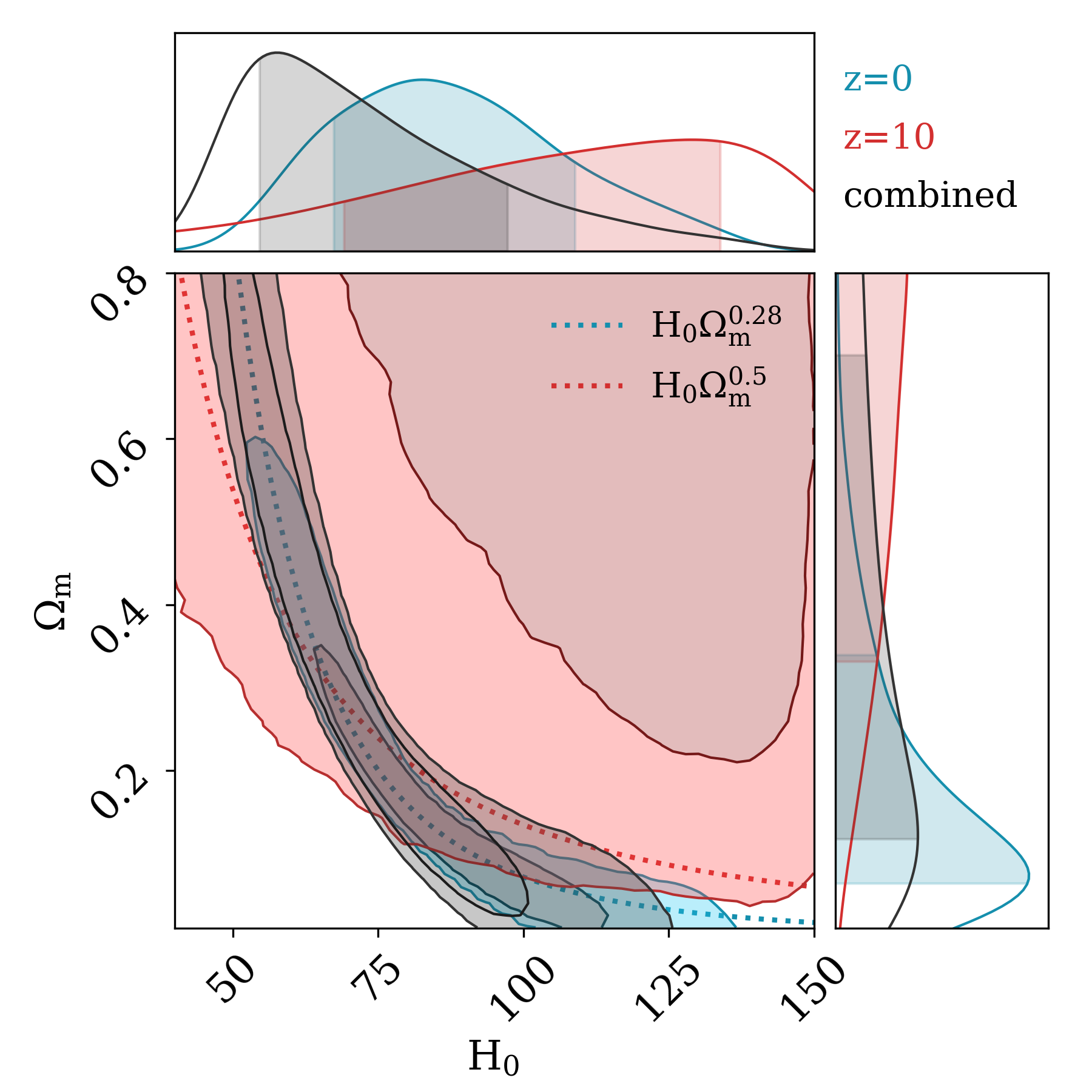}
         \caption{}
         \label{fig:ho_om}
     \end{subfigure}
     \begin{subfigure}[b]{0.43\textwidth}
         \centering
         \includegraphics[width=\textwidth]{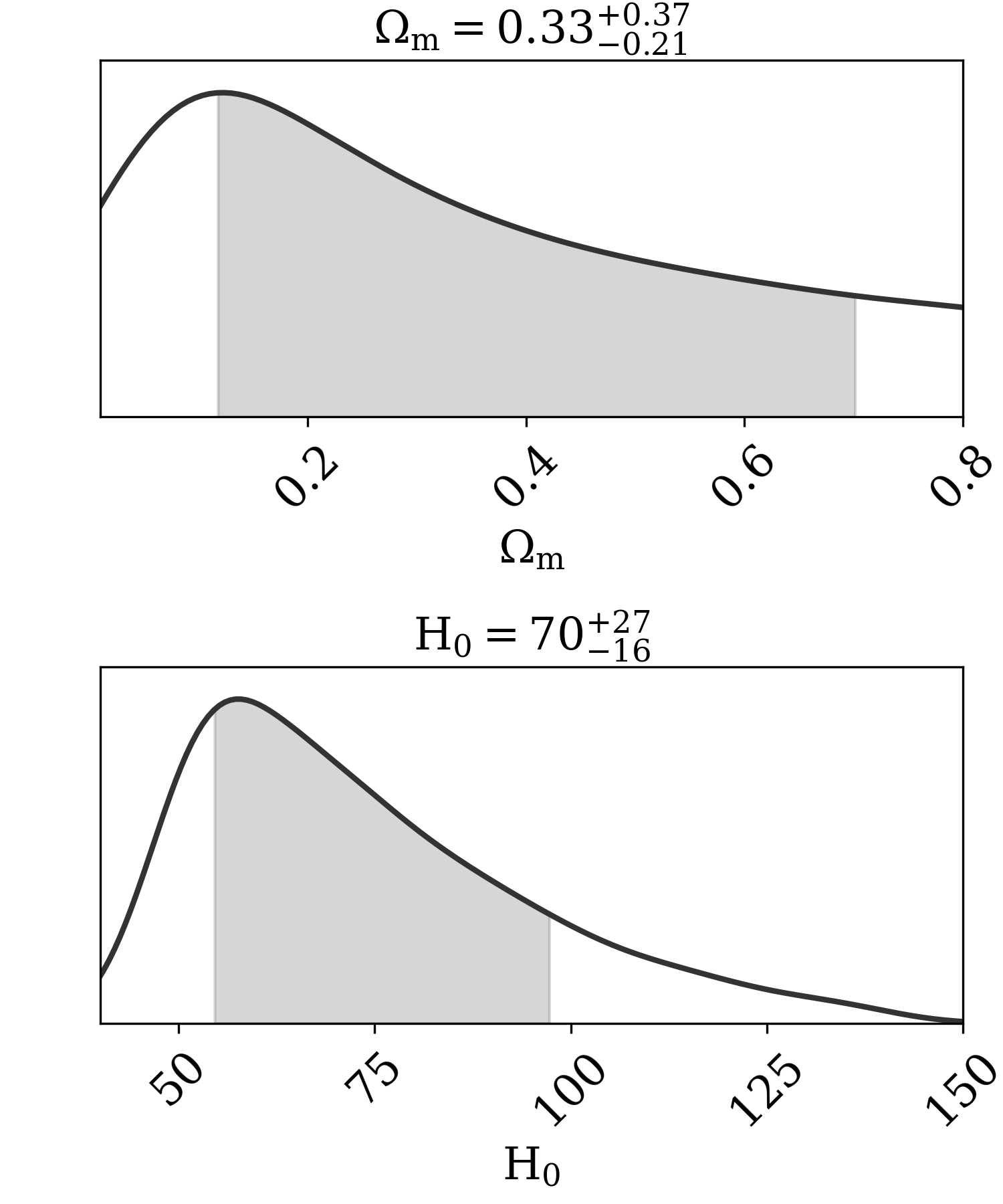}
         \caption{}
         \label{fig:h0_om_1d}
     \end{subfigure}
     \caption{Cosmological constraints from current data. \textit{Left}: Results of the age-z fit in the $H_0 - \Omega_m$ plane. Blue contours refer to the fit of local GCs only, while the red results from adding the ages for the $z=9.625$ cosmic gems, starting to break the $H_0 - \Omega_m$ degeneracy. With dotted lines, the theoretical degeneracy lines derived in Sect. \ref{sec:4cosmo_theory} for the local (in blue) and high-redshift (in red) regimes are shown. \textit{Right}: $\Omega_m$ and $H_0$ 1-D posteriors for the combined fit of the local GCs and the Cosmic Gems ($z=9.625$) ages. $H_0$ is expressed in units of $\rm{km\ s^{-1}\ Mpc^{-1}}$.}\label{fig:fit_mock}
\end{figure*}

The stellar ages we measure for lensed GCs are completely independent of any underlying cosmological assumption.
We can therefore use them as direct cosmic clocks, provided we can robustly infer the age of the Universe, $t_U$, from the age of the clusters, $t_{\star}$ \citep[see][and Sec.~\ref{sec:4cosmo_fit_data} below for two ways of doing this]{Valcin2021}. 

First, we illustrate how age measurements can constrain the parameter space in a flat $\Lambda$CDM cosmology.
Second, we show the cosmology constraints that can be obtained from the current data, combining the Gems at $z=9.625$ and the local globular clusters at $z=0$ \citep{valcin_age_2026}. Finally, we forecast by testing the limits of the method and its resilience under realistic observation.

\subsection{Approximate scalings in $\Lambda$CDM}
\label{sec:4cosmo_theory}
In a flat $\Lambda$CDM cosmological model, the age of the Universe as a function of redshift, $t(z)$, can be expressed as:
\begin{equation}\label{eq:t(z)}
    t(z) = \frac{1}{H_0}\int_z^{\infty}\frac{1}{(1+z')E(z')} dz',
\end{equation}
where \textit{E(z)}, defined as \textit{H(z)}/$H_0$, is a function of redshift and matter density parameter (at $z=0$) $\Omega_m$:
\begin{equation}\label{eq:5H0LCDM}
    E(z)=\sqrt{\Omega_m(1+z)^3+(1-\Omega_m)}
\end{equation}
%5
and we focus on epochs where radiation is negligible.
The age of the Universe, $t(z=0)=t_0$, at the present time (Eq.~\ref{eq:t(z)}) becomes:
\begin{equation}
    t_0 = \frac{F(\Omega_m)}{H_0}, \quad F(\Omega_m) = \frac{2}{3\sqrt{1-\Omega_m}} \rm{sinh}^{-1} \sqrt{\frac{1-\Omega_m}{\Omega_m}}. 
\end{equation}
After applying natural logarithms, the total differential can be written as:
\begin{equation}
    d (\ln t_0) = -d(\ln H_0) + \frac{d \ln{F(\Omega_m)}}{d \ln{\Omega_m}} d\ln{\Omega_m}.
\end{equation}
Around a fiducial $\Omega_m \sim 0.3$, the ratio $d \ln{F}/d \ln{\Omega_m} \sim -0.28$. With this approximation, the local $H_0 - \Omega_m$ degeneracy direction at fixed age follows from:
\begin{equation}\label{eq:deg_z=0}
    d(\ln{H_0}) \simeq -0.28\: d(\ln{\Omega_m}) \quad \Rightarrow \quad \Omega_m^{0.28}H_0 \simeq \rm{const}. 
\end{equation}
When measuring the local age of the Universe, then, we are primarily constraining this relation between $H_0$ and $\Omega_m$ rather than either parameter individually. For this reason, $H_0$ constraints cannot be derived directly from the age of the oldest local objects, but require an external $\Omega_m$ prior to break the degeneracy \citep[see, e.g.,][]{Cimatti2023,tomasetti_globular_2025,tomasetti_oldest_2026}.

However, moving to higher redshift, the Universe becomes increasingly matter-dominated, and the contribution from the cosmological constant in Eq. \ref{eq:5H0LCDM} becomes negligible. For $z\gg 1$, $E(z)$ simplifies to:
\begin{equation}
    E(z) = \sqrt{\Omega_m}(1+z)^{3/2}.
\end{equation}
This implies that:
\begin{equation}
    t(z) = \frac{1}{H_0 \sqrt{\Omega_m}} \int_z^\infty (1+z')^{-5/2} dz' =\frac{2}{3H_0\sqrt{\Omega_m}} (1+z)^{-3/2},
\end{equation}
which, at fixed redshift, becomes:
\begin{equation}\label{eq:deg_z>>1}
    H_0 \sqrt{\Omega_m} \simeq \rm{const}.
\end{equation}
%
%The picture emerging from Eq. \ref{eq:deg_z=0} and Eq. \ref{eq:deg_z>>1} is that
Moving from the local Universe at $z \sim 0$, to higher redshifts ($z \gg 1$), the parameter degeneracy effectively rotates in the $H_0 - \Omega_m$ plane, shifting from a power-law dependency of the form $\Omega_m^{0.28}$ to a shallower one following $\Omega_m^{0.50}$. Around a fiducial $\Omega_m\sim0.3$, in particular, the degeneracy scales with $\Omega_m^{0.48}$ by $z\sim2$ and reaches the form in Eq.~\ref{eq:deg_z>>1} at the 1\% level by $z\sim3$.

As a consequence, several measurements of the age of the Universe at different redshifts can effectively break the $H_0 - \Omega_m$ degeneracy, and the cosmic clocks approach can become a fully independent and self-consistent cosmological probe.

In this context, the oldest lensed stellar clusters, potentially representing the progenitors of local globular clusters, represent the optimal tracer to constrain the age of the Universe up to high redshift.

First, being the earliest collapsed structures, their formation epoch drastically narrows down to a limited redshift window, close to the Big Bang. Second, they are among the closest approximations to simple stellar populations. Their limited internal age spreads reduce parameter degeneracies and simplify SED modelling. If young and intermediate-age massive clusters in the local Universe are representative analogues of GCs at formation, their star-formation histories may have lasted only a few Myr \citep{martocchia_search_2018}. More generally, even classical multiple-population formation scenarios usually restrict any extended star formation to at most a few hundred Myr \citep[e.g.,][]{Bastian2018,Gratton2019}. Their ages are therefore substantially better constrained than those of composite stellar systems. Third, they are among the few approximately simple stellar populations that can be detected as individual systems at such extreme distances, making them exceptionally valuable clocks for the early Universe.

Importantly, high-redshift clusters do not need to be the direct progenitors of today's GCs to serve as cosmic clocks. Even in the hypothesis that many of them eventually disrupt, they remain key probes of the high-z Universe. What matters is that they are coeval star clusters, born during the same early cosmic epoch as surviving local GCs, and with similar properties, allowing us to model both populations under the same physical assumptions.

\begin{figure*}[t] 
     \centering
     \begin{subfigure}[b]{0.49\textwidth}
         \centering
         \includegraphics[width=\textwidth]{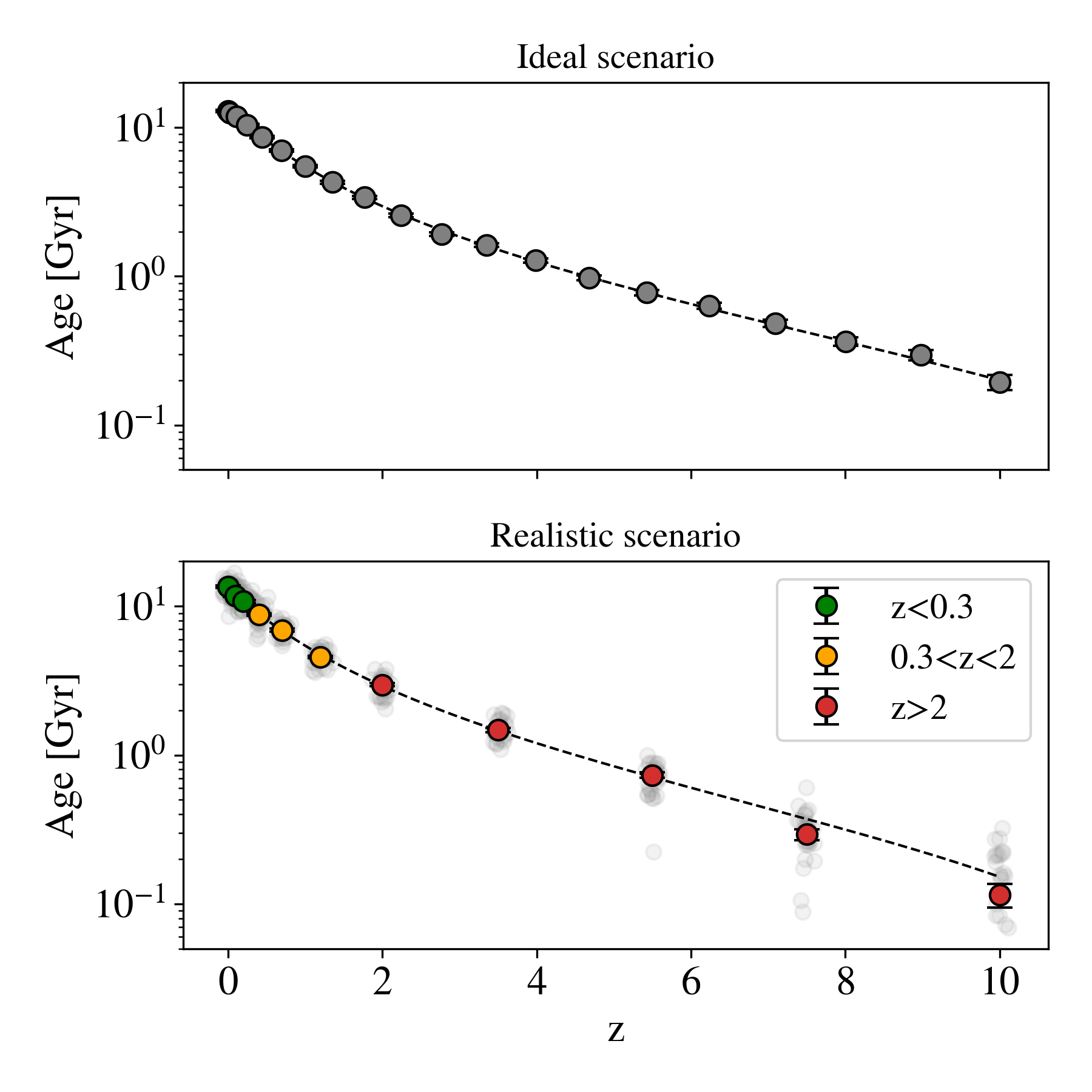}
         \caption{}
         \label{fig:age_mock}
     \end{subfigure}
     %\hfill 
     \begin{subfigure}[b]{0.47\textwidth}
         \centering
         \includegraphics[width=\textwidth]{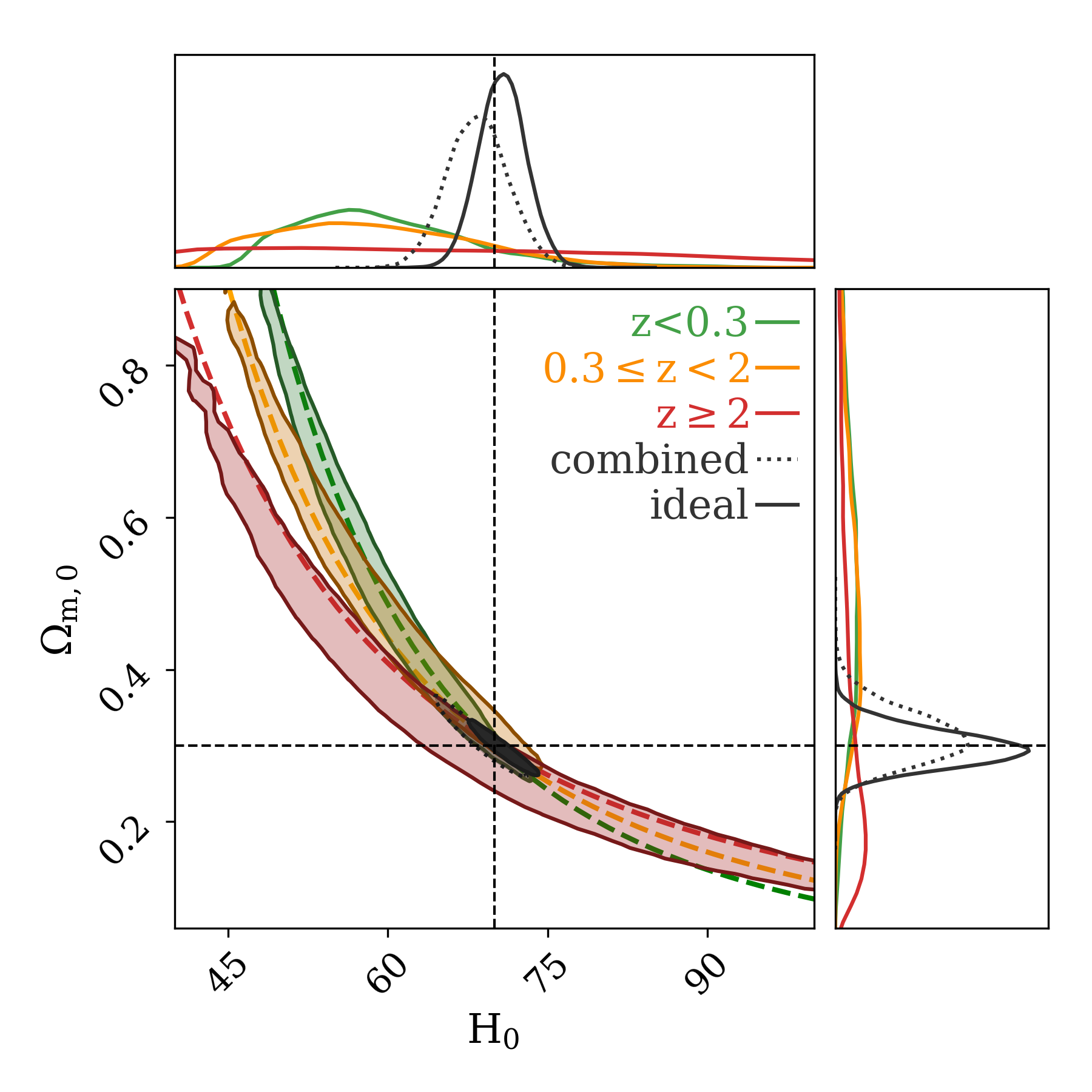}
         \caption{}
         \label{fig:ho_om_mock}
     \end{subfigure}
     \caption{Cosmological constraints from mock samples. \textit{Left:} Mock age -- redshift relation based on a flat $\Lambda$CDM cosmology, for both the ideal (top) and realistic (bottom) scenarios (see Sect. \ref{sec:4cosmo_fit_mocks}). In the latter, we show the whole population (N $\sim 300$) in grey while the mean values at each redshift are shown in different colours depending on the redshift range: low-z (green), mid-z (yellow), and high-z (red). \textit{Right:} 1-$\sigma$ contours in the $\Omega_m-H_0$ plane obtained by fitting the age -- redshift trend within the three redshift ranges of the realistic scenario, either taken separately (with corresponding colours) or jointly (dotted black contour). In black are shown the contours resulting from the fit in the ideal scenario. The dashed lines in the background show the $H_0 - \Omega_m$ degeneracy rotation described in Sect.~\ref{sec:4cosmo_theory}, from Eq.~\ref{eq:deg_z=0} (in green) to Eq.~\ref{eq:deg_z>>1} (in red).}
\end{figure*}

\subsection{Reconstructing $t(z)$ up to $z=10$}
\label{sec:4cosmo_fit_data}

We consider the newly derived age measurements for the Gems at $z \approx 9.6$ alongside the local GC population. For the local GC, we consider the state-of-the-art compilation from \citet{valcin_age_2025,valcin_age_2026}. 

Using this combined dataset, we perform a joint fit to the cosmic time relation expressed in Eq. \ref{eq:t(z)}, adapting the extremes of the integral as follows. %to each case. 
In each case, the lower bound corresponds to the (known) redshift of observation, while the upper bound represents the (unknown) redshift of formation of each population. Consequently, instead of adopting a prior distribution for formation redshift, this approach, conservatively, introduces $z_{f}$ as an additional free parameter in our cosmological fitting procedure, alongside the Hubble constant ($H_0$) and the matter density parameter ($\Omega_m$).
For this reason, these old systems provide the most stringent cosmological constraints, as their proximity to the Big Bang significantly narrows the allowed parameter space for $z_{f}$. The Gems, in particular, are observed at such an early epoch that they effectively push the formation limit toward the very beginning of the reionization era.

In our joint analysis, we fit these two populations simultaneously while treating their formation redshifts ($z_{f,\text{local}}$ and $z_{f,\text{Gems}}$) as independent parameters. This approach ensures that the formation histories of clusters at $z \approx 0$ and $z \approx 10$ are reconstructed independently. For the local GCs, we considered the single, but very precise, constraint produced by \citet{valcin_age_2026}: $t_{\rm{GC}} = 13.61 \pm 0.25\: \rm{(stat)} \pm 0.23\: \rm{(sys)}$. For the Gems, we include the oldest half of our sample, $t_{\rm{Gems}} = 102 \pm 63$ Myr, obtained as described in Sect. \ref{sec:Gems_ages}.

We performed a Monte Carlo Markov Chain (MCMC) analysis using the affine-invariant ensemble sampler \texttt{emcee} \citep{ForemanMackey2013}, employing 250 walkers with 1,000 iterations each and a burn-in phase of 300 steps to ensure convergence. We adopted non-informative uniform priors over the following ranges: $H_0 \in [0, 150]$ km s$^{-1}$ Mpc$^{-1}$, $\Omega_m \in [0.01, 0.99]$, and $z_f \in [9.7, 30]$ for both the local and high-redshift sample, ensuring that the joint constraints are driven exclusively by the data. The resulting chains were then marginalized over the independent formation redshifts to extract the final cosmological parameters.

In Figure~\ref{fig:ho_om}, we report the constraints in the $H_0-\Omega_m$ plane obtained separately from the local GCs ages (blue contours) and the Cosmic Gems ages (red contours). Dotted lines in the same colours represent the theoretical degeneracy lines recovered in the previous section. On their own, given the limited statistics and limited redshift samples, individual constraints are heavily dominated by the parameter degeneracy. In the case of the Gems, in particular, the age constraint remains relatively young and with low precision; as a consequence, the corresponding confidence limits are basically just excluding part of the parameter space, removing the low-$\Omega_m$ region. This means that the Gems data effectively put only an upper limit to the age of the Universe at $z\sim 10$. On the other hand, a lower age limit is not captured because a very young Universe, predicted by high values of $\Omega_m$ and $H_0$, cannot be ruled out with such a young and highly uncertain age measurement.

What is interesting to observe, however, is the complementarity of the constraints coming from these two redshift regimes: the local GCs track the full $H_0-\Omega_m$ degeneracy, essentially setting an upper limit on $\Omega_m$; the high-$z$ ages, instead, penalize very low $\Omega_m$ values. This means that including high-z measurements, despite the limited precision, effectively induces the rotation of the degeneracy ellipses discussed before (indicated in Fig.~\ref{fig:ho_om} by the dashed lines), successfully excluding the high-$\Omega_m$, high-$H_0$ region in the joint fit, and yielding $H_0=70^{+27}_{-16}\ \rm{km\ s^{-1}\ Mpc^{-1}}$ and $\Omega_m=0.33^{+0.37}_{-0.21}$, where each value represents the central credible interval of the marginalised distributions.

\section{Forecasting future constraints}
\label{sec:4cosmo_fit_mocks}

The methodology developed demonstrates the power of high-redshift stellar clusters to serve as cosmic clocks. While current results are limited by the small sample size and poor redshift coverage, the number of detected lensed and unlensed high-redshift sources is growing rapidly.
A recent study \citep{claeyssens_first_2026} identifies more than 200 lensed stellar clusters up to $z\simeq8$ just within the field of the Abell S1063 galaxy cluster, leveraging the ultra-deep imaging provided by the GLIMPSE public survey \citep{atek_jwsts_2025}. This massive ensemble joins a growing list of lensed systems discovered across the high-redshift Universe. Notable examples include the Sunrise Arc \citep[$z=6$,][]{vanzella_jwstnircam_2023}, the Cosmic Grapes \citep[$z=6.1$,][]{fujimoto_primordial_2025}, the Cosmic Spear \citep[$z=6.2$][]{abdurrouf_spatially_2025}, the Firefly Sparkle \citep[$z=8.3$,][]{Mowla2024}, and, even at $z>10$, the Bullet Arc \citep[$z=11.1$][]{bradac_star_2025} and the Misty Moons \citep[$z=11-12$][]{nakane_venus_2025}. Another promising avenue for densely populating the age-redshift relation at $z<1$ relies on the study of redshifted, but not lensed, globular clusters within galaxy clusters. A notable example is that of MACS0417.5-1154 at $z\sim 0.44$, where \citet{harris_jwst_2025} identified a population of $\sim 10^5$ GCs. 

Not all these sources will turn out to be optimal for this analysis, however, with such large samples of stellar clusters, also the population of long-lived objects will naturally increase, enabling a robust mapping of the oldest stellar populations from the local Universe up to very high redshift. 
With JWST, lensed clusters are detected in almost every lensing galaxy cluster, while Euclid's wide-field survey is expected to uncover thousands of new lensing systems. Looking ahead, the upcoming ELTs will make the observation of these compact systems a routine task. In this context, it is crucial to quantify the data quality and the sample size required to obtain cosmologically competitive results. In the following, we explore how the precision on $H_0$ and $\Omega_m$ scales with the number of high-redshift measurements and their associated age uncertainties. Naturally, the resulting forecasted errors represent a statistical error-floor in the absence of significant systematics.

\subsection{The ideal scenario: dense sampling of coeval populations}

To optimize the constraining power of the method, we first consider an ideal mock dataset where the formation redshift is fixed exactly to $z_f = 15$ for all objects. Furthermore, the data points are distributed across the cosmic timeline using a quadratic spacing that samples 20 distinct intervals from $z = 0$ up to $z = 10$. This configuration provides a very dense and smooth distribution at low and intermediate redshifts, allowing us to evaluate the performance of our methodology in the absence of statistical scatter on $z_f$. 

In generating this mock sample, the age at each redshift is calculated via Eq. \ref{eq:t(z)} assuming $H_0=70\: \rm{\kmsmpc}$ and $\Omega_m = 0.3$, by computing the integral between $z_{\text{obs}}$ and its assigned $z_f$. For each redshift bin, we assume the age anchor is derived from a sample of $N=30$ independent clusters. Each individual cluster is assigned an error of 10\% of its age, with a floor of 100 Myr to prevent unrealistically small error bars at young ages. The final uncertainty for each redshift bin is then scaled down by $\sqrt{N}$. This is shown in the top left panel of fig.~\ref{fig:ho_om_mock}.

We analyse these mocks in the same way as the real data by performing MCMC fitting to the mock age-redshift relation as described in Sect.~\ref{sec:4cosmo_fit_data}; the resulting posteriors are shown in black in Fig.~\ref{fig:ho_om_mock}. This provides the following marginalized constraints:
\begin{equation*}
    \begin{aligned}
        H_0      &= 70.7^{+2.2}_{-2.1} \text{ km s}^{-1} \text{ Mpc}^{-1} \quad &(\text{truth: } 70 \rm{\kmsmpc}) \\
        \Omega_m &= 0.29^{+0.02}_{-0.02} \quad &(\text{truth: } 0.30) \\
        z_f      &= 15.3^{+0.6}_{-0.5} \quad & (\text{truth: } 15)
    \end{aligned},
\end{equation*}
where the uncertainties represent the 16th and 84th percentiles of the marginalized posterior distributions.

The recovered parameters and associated error bars from this ideal scenario illustrate that our methodology has the potential to constrain cosmology in a very competitive way. A precision of approximately 3\% on $H_0$ and 7\% on $\Omega_m$ makes this approach comparable with standard alternative probes, such as Baryon Acoustic Oscillations (BAO) and cosmic shear. But even more importantly, our method achieves this level of precision without requiring external priors or calibrations, as is the case for the CMB or for local distance ladders.

\subsection{The realistic scenario}
A more realistic simulation setup considers a mock dataset where the redshift of formation is no longer fixed, but accounts for the intrinsic scatter of a physical sample of stellar clusters. To build it, we start from the assumption that these very pristine objects should follow a specific cosmic SFH, with a peak closer to the reionization epoch and a progressively lower probability of formation as we move towards $z_f \sim 20-30$. \citet{madau_cosmic_2014}, for instance, report a rising phase in the cosmic SFH scaling as $\psi(z) \propto (1 + z)^{-2.9}$ at $3 \lesssim z \lesssim 8$. Here, we follow the same functional form, but since our simulation probes the very young Universe ($z > 10$), we adopt a steeper decline so that the SFH effectively drops to zero by $z = 30$. Our probability density function is thus defined as:
\begin{equation}\label{eq:cosmic_SFH}
    f(z_f) \propto (1 + z_f)^{-5} \quad 10 \le z_f \le 30,
\end{equation}
where the steep exponent of $-5$ accounts for the rapid suppression of star formation in the primordial Universe, and where the distribution is strictly bounded between a minimum threshold of $z = 10$ and a maximum cutoff of $z = 30$.

Drawing simulated ages directly from this distribution, which has a large intrinsic scatter, would introduce a fundamental methodological challenge: these structures could have formed at vastly different epochs, meaning that we could not use a single, global $z_f$ parameter to fit the age constraints across the entire redshift range. Indeed, the relevant stochastic bias would be too large, as any individual cluster might be an outlier formed much earlier or later than the average population.

\begin{figure}
    \centering
    \includegraphics[width=\linewidth]{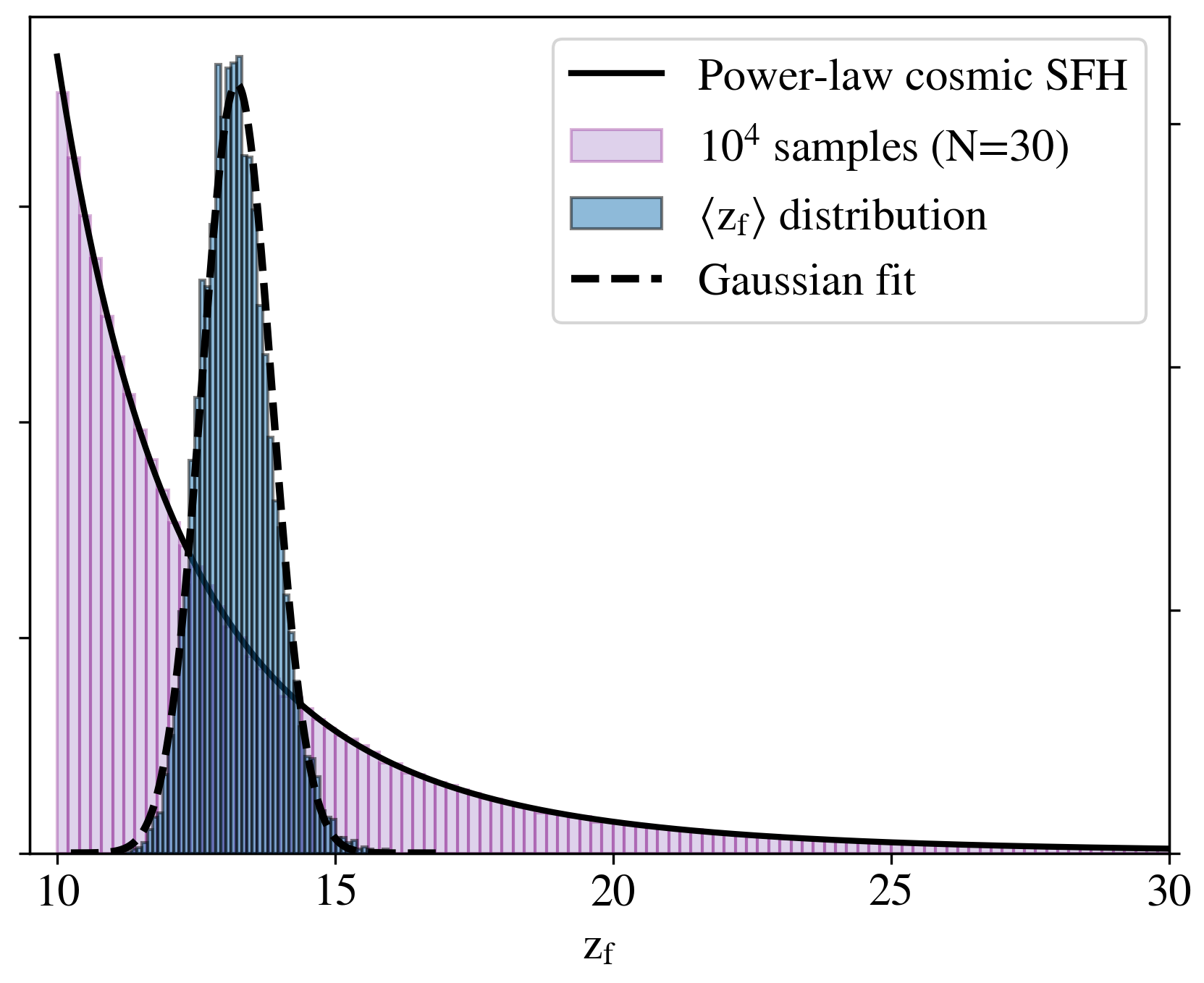}
    \caption{Distribution of the formation redshift, $z_f$, assumed in the realistic simulation. In purple, we show the $10^4$ realizations of samples with 30 GCs each, drawn randomly from the power-law assumed as a cosmic SFH. In blue, the resulting Gaussian distribution of the mean values of each sample drawn. In black, the resulting Gaussian fit, with $\langle z_f\rangle = 13.3$ and $\sigma = 0.6$.}
    \label{fig:CLT}
\end{figure}

Instead, assuming that each redshift is populated by a statistically significant ensemble, like a sample of 30 objects per redshift bin, allows us to exploit the Central Limit Theorem: averaging over a large sample turns the heavily skewed, power-law distribution of the individual clusters into a stable, symmetric Gaussian distribution for the bin. To compute the mean and standard deviation of the Gaussian corresponding to our cosmic SFH, we generate $10^4$ samples, composed of 30 clusters each, drawing $z_f$ from Eq.~\ref{eq:cosmic_SFH}. This results in $\langle z_f\rangle= 13.3 \pm 0.6$. Fig.~\ref{fig:CLT} shows the randomly generated samples from Eq.~\ref{eq:cosmic_SFH} and the resulting Gaussian distribution for $z_f$.\\

In this simulation, the formation redshift assumed for each redshift bin was thus extracted from a Gaussian distribution with $\mu=13.3$ and $\sigma=0.6$. We can also observe that while the true cosmic SFH may differ, regardless of its shape, the resulting distribution of these averages universally converges toward a Gaussian with a significantly reduced standard deviation. In addition, to meet realistic observational capabilities, we adopt a sparser redshift coverage with respect to the ideal case, spanning 11 bins from the local Universe up to $z = 10$, as shown in the bottom panel of Fig. \ref{fig:age_mock}. Age measurements and associated errors were derived at each redshift following the procedure described in the previous section. Repeating the MCMC analysis, the cosmological constraints in this realistic scenario result in:

\begin{equation*}
    \begin{aligned}
        H_0      &= 68.4^{+2.8}_{-2.8} \text{ km s}^{-1} \text{ Mpc}^{-1} \quad &(\text{truth: } 70 \:\rm{\kmsmpc}) \\
        \Omega_m &= 0.31^{+0.04}_{-0.03} \quad &(\text{truth: } 0.30) \\
        z_f      &= 12.3^{+0.4}_{-0.4} \quad & (\text{truth: } 13.3\pm0.6)
    \end{aligned}
\end{equation*}

and are shown with black dotted lines in Fig.~\ref{fig:ho_om_mock}. In the same figure, we show with bright colours the constraints coming from fitting the same mock dataset, separately for the three redshift intervals, clearly illustrating the rotation of the degeneracy contours from $z=0$ to $z=10$. 
This shows that, even in this realistic scenario, the analysis successfully recovers the input parameters of the physical mock catalog within $1\sigma$, with no significant bias in $H_0$ and $\Omega_m$. The formation redshift, instead, is underestimated at $1.4\sigma$ level. This effect likely arises from the non-linearity of the redshift-to-age conversion. When assuming a Gaussian distribution in $z_f$ space, the non-linear mapping distorts this into an asymmetric, skewed distribution in age, favouring younger values. Since we associate symmetric Gaussian error bars to the age measurements when generating the mock, the MCMC struggles to capture this inherent asymmetry, pulling down the median $z_f$. Importantly, however, this systematic shift does not introduce significant biases in the recovery the cosmological parameters of interest.

The precision achieved remains remarkably high, yielding a relative uncertainty of $4\%$ on $H_0$ and $11\%$ on $\Omega_m$. This is a highly encouraging result: despite the observational limitation given by the sparse 11-bin redshift sampling and the physical scatter on the formation redshift, the constraints suffer only a minor degradation compared to the ideal case. 

The forecasted precision within the flat $\Lambda$CDM framework is comparable to the accuracy obtained from other well-established surveys, like the Dark Energy Survey final dataset \citep[DES Y6][]{abbott_dark_2026} when leveraging multiple probes. Combining BAO constraints from $\sim$1.6 million galaxies, Hubble diagram measurements from $\sim1800$ SNIa, and modern Big Bang Nucleosynthesis priors \citep{schoneberg_2024_2024}, an accuracy of roughly $5\%$ on both $H_0$ and $\Omega_m$ is achieved.

The fact that a single, self-consistent cosmic probe, leveraging a few hundred high-redshift star clusters, can approach the constraining power of other multi-probe frameworks clearly highlights the great potential of this method, particularly as future dedicated surveys are set to drastically increase the available sample size.

While our current statistical constraints are encouraging, a detailed analysis of systematic effects is mandatory, as they could play a major role in our final cosmological results. In particular, the impact of stellar models needs to be evaluated: for instance, models with different $\alpha$-enhancement \citep{park_-mc_2025} or including binary stellar evolution \citep{eldridge_binary_2017}. We present a preliminary assessment of the latter in Appendix \ref{sec:App_comparison}, where we compare our results with literature estimates derived from binary-star models. At present, the differences introduced by binary physics remain well within our statistical uncertainties. Once a larger, statistically significant sample of these sources becomes available, it will be necessary to rigorously quantify the systematic budget to evaluate the effective applicability and accuracy of high-redshift star clusters as cosmological probes.

\section{Conclusions}\label{sec:5CONCLUSIONS}

In this work, we presented a novel cosmological application of high-redshift stellar clusters as cosmic clocks, launching the \textsc{CORALS} project. Specifically, we extended the methodology originally proposed in \citet{tomasetti_time_2025} to the extreme redshift regime, leveraging the highly magnified Cosmic Gems arc at $z=9.625$ \citep{adamo_bound_2024,messa_jwst_2026}. 

First, we obtained deconvolution photometry with \texttt{STARRED}, successfully identifying 20 distinct point sources, which correspond to 10 physically unique lensed stellar clusters. This effectively doubles the number of individual cluster SEDs currently available in the literature for this system.

We performed SED fitting using \texttt{BAGPIPES} without imposing a cosmological prior on the age of the universe, allowing us to extract unbiased stellar ages that can be used as cosmological proxies. To ensure the robustness of our results, we tested three distinct SFHs, finding that the resulting age estimates are fully consistent across all three models within their uncertainties. Adopting the exponentially declining SFH as our benchmark, we constrain the combined age of the oldest half of the sample to $t_\star = 102 \pm 63\,\mathrm{Myr}$.

We then explored the cosmological implications of going to high-$z$, demonstrating analytically that the $H_0-\Omega_m$ degeneracy rotates from $z = 0$ out to the high-redshift regime ($z \gg 1$). As a practical test of this mechanism, we combined the ages of local GCs with our high-$z$ Cosmic Gems measurements within a joint fit, leaving $H_0$, $\Omega_m$, $z_{f,\mathrm{local}}$, and $z_{f,\mathrm{gems}}$ as free parameters. This resulted in $H_0 = 70^{+27}_{-16}\,\mathrm{km}\,\mathrm{s}^{-1}\,\mathrm{Mpc}^{-1}$ and $\Omega_m = 0.33^{+0.37}_{-0.21}$. While the current sample size does not allow for precise cosmological constraints, this analysis observationally confirms the distinct dependencies in place at $z=0$ versus $z \sim 10$, successfully narrowing the allowed parameter space by rotating the degeneracy ellipses.

Finally, we investigated the potential of this technique through cosmological forecasts. We simulated a realistic future sample of approximately 300 lensed stellar clusters distributed across $z=0$ to $z=10$, accounting for realistic uncertainties in both the observational age measurements and the formation redshifts ($z_f$). Our forecasts demonstrate that such a sample can place competitive constraints, achieving a statistical precision of $4\%$ on $H_0$ and $11\%$ on $\Omega_m$. This prospect is particularly timely, as the synergy between current JWST observations, Euclid's ongoing wide-field discoveries, and the upcoming ELTs will turn this simulation into a realistic observational sample in the near future.

\begin{acknowledgements}
The authors warmly thank Matteo Messa and Eros Vanzella for the useful and constructive feedback on this work. ET, MMo, and AC acknowledge support from the grant ASI n. 2024-10-HH.0 “Attività scientifiche per la missione Euclid – fase E”. ET acknowledges the support from COST Action CA21136 – “Addressing observational tensions in cosmology with systematics and fundamental physics (CosmoVerse)”, supported by COST (European Cooperation in Science and Technology). MMo acknowledges support from the MIUR PRIN 2022 grant "Optimizing the extraction of cosmological information from Large Scale Structure analysis in view of the next large spectroscopic surveys" (grant $2022NY2ZRS001$). Funding for the work of RJ, LV was partially provided by project PID2022-141125NB-I00; FC is in part supported by the Agencia Estatal de Investigación (AEI), Ministerio de Ciencia, Innovación y Universidades, Spain, under the project PID2024‑155455NB‑I00, within the 2024 Call for Knowledge Generation Projects. RJ, FC and LV acknowledge  the“Center of Excellence Maria de Maeztu 2020-2023” award to the ICCUB (CEX2019- 000918-M) funded by MCIN/AEI/10.13039/501100011033. MMi acknowledges support by the SNSF (Swiss National Science Foundation) through the return CH grant P5R5PT\_225598 and Ambizione grant PZ00P2\_223738. 
    
\end{acknowledgements}

\bibliographystyle{aa}
\bibliography{references}

\newpage
\begin{appendix}

\section{Gems deconvolution photometry with \texttt{STARRED}}\label{sec:app_deconv}

We perform a dedicated deconvolution procedure on the JWST imaging, across all eight NIRCam bands: F090W, F115W, F150W, F200W, F277W, F356W, F410M, and F444W (GO 4212, P.I. Bradley). We utilized the standard data products from the MAST archive, where the SW images are sampled at $0.031~\arcsec$ per pixel and LW images have a pixel scale of $0.063~\arcsec$, corresponding to the nominal instrument resolution.

Given the presence of blended components within the Cosmic Gems Arc, and for consistency with the Sparkler photometry, we employ \texttt{STARRED} \citep{millon_image_2024, Michalewicz2023}, the two-channel deconvolution pipeline already used to obtain the photometry of the GCs in the Sparkler, specifically designed to separate compact, point-like sources from underlying diffuse structures. This separation is achieved by decomposing the image into the Starlet domain, which allows the code to capture extended features across different spatial frequencies.\\
We first reconstruct the narrow point spread function (PSF) for each band using six bright stars near the Gems, then use the identified PSFs to perform a deconvolution of the data, improving the sampling by a factor of 2. This oversampling yields a final pixel size of $0.016$ \arcsec for the SW bands and $0.031$ \arcsec for the LW bands, effectively doubling the spatial resolution relative to their respective native detector scales. 

As discussed in the main text, the arc consists of two mirrored images of the clusters, yielding two distinct SEDs for each source. Previous work by \citetalias{adamo_bound_2024} focused on the five primary clusters (ABCDE) and \citetalias{messa_jwst_2026} identified four additional faint point sources in the tails of the arc. While these are consistent with being star clusters, given the lower magnification they undergo, they are also compatible with slightly more extended stellar structures ($R_{\rm{eff}}\geq20$ pc). 
\texttt{STARRED} allows us to deblend and extract the SEDs for both central and tail sources, and resolve a sixth pair of sources (F1--F2) within the innermost region. Contrary to the tail sources, this innermost region undergoes extreme magnification. As a result, the observed point-like nature of this detection would imply an ultra-compact, sub-parsec, intrinsic radius according to the current lens model.

\begin{figure}
    \centering
    \includegraphics[width=0.9\linewidth]{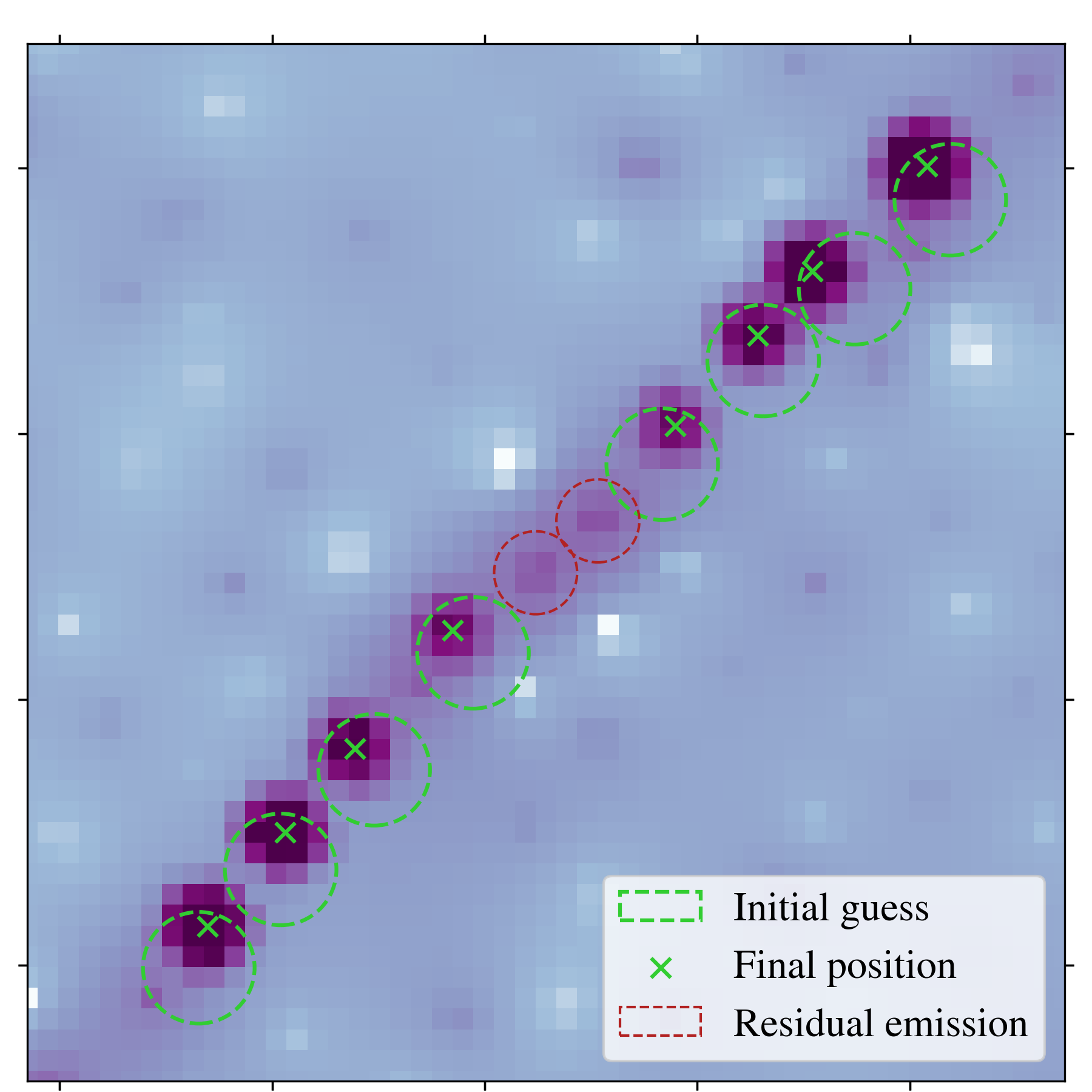}
    \caption{Result of the first round of deconvolution. The image is zoomed into the innermost region, excluding sources A1–A2 and the tails. Green dashed circles represent the initial positions provided for point-source detection in \texttt{STARRED}; note that only 8 point sources (corresponding to B1-E1 and B2-E2) are provided in this phase. Green crosses indicate the final best-fit positions after deconvolution. Red dashed circles highlight the two distinct residuals emerging in the innermost region, which motivate the search for an additional pair of point sources.}
    \label{fig:first_deconv}
\end{figure}

This additional point emerges naturally from the iterative deconvolution process, which was initiated by providing \texttt{STARRED} only with the positions of the five central sources (A–E) as an initial guess for point-source detection. Without any other prior information, this preliminary deconvolution run revealed a highly prominent, structured residual in the central region of the image, clearly highlighted by the red dashed circle in Fig. \ref{fig:first_deconv}, which shows the result of the first deconvolution run. This residual clearly splits into two distinct, point-like components.
Note that while \citetalias{messa_jwst_2026} labeled the first of the external tail sources as 'F', we reserve the label 'F' for this newly resolved central source, consequently naming the four external tail sources as G--J.

All 20 point sources are robustly identified in both the F150W and F200W images. Due to the lower resolution in the other long-wavelength bands, which prevents independent identification, we adopt the coordinates derived from these higher-resolution filters as narrow priors for the deconvolution in the LW bands.

\section{Measuring ages}\label{sec:app_FSF}

For the SED modeling within \texttt{BAGPIPES} \citep{Carnall2018}, we adopted the 2016 version of the \citet{Bruzual2003} stellar population synthesis models \citep[see also][]{Chevallard2016}, assuming a \citet{kroupa_variation_2001} Initial Mass Function (IMF). 

We test three different SFHs, defined as follows. The single burst model represents an instantaneous star formation event at a single lookback time. The exponentially declining model follows a $\tau$ parametrisation where $\mathrm{SFR}(t) \propto \exp(-t/\tau)$. The delayed exponentially declining (DED) model follows the smooth functional form $\mathrm{SFR}(t) \propto (t-T_0)\exp(-(t-T_0)/\tau)$. For both the exponential and DED models, the e-folding timescale $\tau$ is governed by a uniform prior ranging from 0 to 0.1 Gyr. Age, instead, is allowed to span up to 5 Gyr, which is almost ten times the age of the Universe at $z=9.625$ in a vanilla $\Lambda$CDM, $\sim 470$ Myr.

Regarding the remaining parameters, which were kept identical across all runs, dust attenuation was included by adopting a \citet{Calzetti2000} attenuation curve. The V-band extinction, $A_V$, was allowed to vary freely under a uniform prior from 0 to 4 mag. Metallicities are allowed in the $[Z/H]$ range from $-4$ to $0.1$ dex.

In Table \ref{tab:sed_results} we report the main physical properties of all sources derived from each configuration. Masses are corrected for magnification based on the values published in \citetalias{messa_jwst_2026}. In particular, the magnification factors used here are approximations derived by averaging the values obtained from their \texttt{glafic} and \texttt{Lenstool} lensing models. Since \texttt{Lenstool} estimates are not available for all sources, we assume them to be approximately $1.5$ times their \texttt{glafic} counterparts (consistent with the offset in Table B1 of \citetalias{messa_jwst_2026}). To account for the lensing uncertainty in the final mass error budget, we scale the lower bound ($16^{\text{th}}$ percentile) of the mass distribution with the maximum magnification ($\mu + \sigma_{\mu}$), and the upper bound ($84^{\text{th}}$ percentile) with the minimum magnification ($\mu - \sigma_{\mu}$).

\begin{figure*}[h!]
    \centering
    \includegraphics[width=\linewidth]{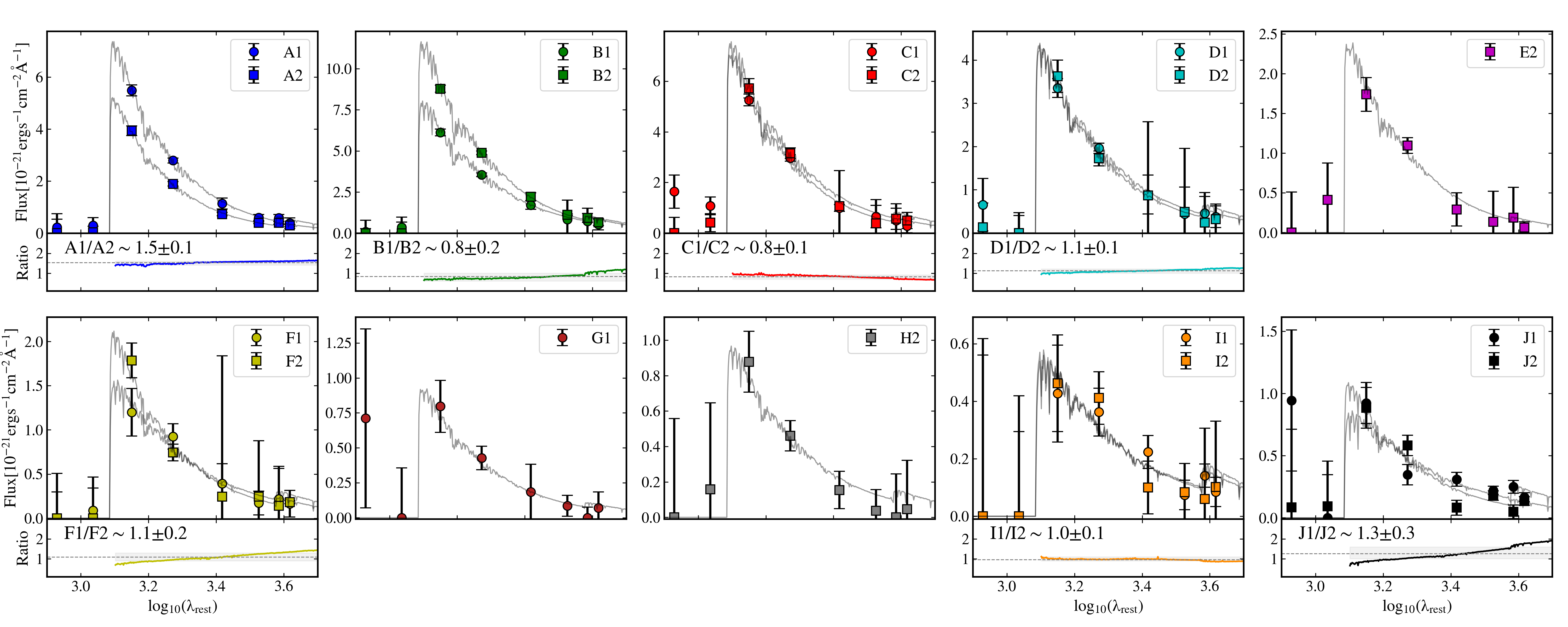}
    \caption{SEDs in all eight filters for the 10 detected clusters. In grey, the spectrum resulting from the fit with an exponentially declining SFH.}
    \label{fig:SEDs}
\end{figure*}

In Fig. \ref{fig:SEDs} we show the SEDs of all 10 sources, with the corresponding fitted spectrum resulting from the DED configuration. Counter-images of the same cluster are shown in the same panel, without correcting for magnification. When both SEDs are available, the lower panel shows the ratio of the two fitted spectra as a function of wavelength, reporting also its mean and standard deviation. This trend is generally flat, compatible with a constant across the whole spectrum, confirming that we are indeed observing counter images of the same source, with identical SEDs up to a multiplicative constant. Additionally, the fact that they are all very close to unity is consistent with the lensing model, suggesting that these mirrored images experience nearly identical magnifications.
\begin{table*}[t]
\centering
\caption{SED fitting results for the three SFH configurations.}
\label{tab:sed_results}
%\small
%\setlength{\tabcolsep}{4pt} % Margine ottimale per farla entrare comodamente nella textwidth
\footnotesize % Più adatto di \small per tabelle a 3 blocchi così dense
\setlength{\tabcolsep}{3.5pt}
\begin{tabular}{l cccc c ccccc c ccccc c c}
\toprule
& \multicolumn{4}{c}{\textbf{Single Burst}} & & \multicolumn{5}{c}{\textbf{Exponentially Declining}} & & \multicolumn{5}{c}{\textbf{Delayed Exponentially Declining}} & & \\
\cmidrule{2-5} \cmidrule{7-11} \cmidrule{13-17}
ID & $t_\star$ & $\frac{Z}{Z_\odot} [\%]$ & $A_V$ & $\log(\frac{M_\star}{M_\odot})$ & & $t_\star$ & $t_{\rm{MW}}$ & $\frac{Z}{Z_\odot} [\%]$ & $A_V$ & $\log(\frac{M_\star}{M_\odot})$ & & $t_\star$ & $t_{\rm{MW}}$ & $\frac{Z}{Z_\odot} [\%]$ & $A_V$ & $\log(\frac{M_\star}{M_\odot}) $ & & $\mu$ \\
\midrule

A1 & $21^{+13}_{-9}$  & $0.5^{+6.1}_{-0.4}$  & $0.3^{+0.1}_{-0.2}$ & $6.9^{+0.3}_{-0.4}$ & & $61^{+51}_{-36}$  &    $35^{+37}_{-22}$ & $0.3^{+5.9}_{-0.3}$  & $0.3^{+0.1}_{-0.2}$ & $6.9^{+0.3}_{-0.4}$ & & $92^{+76}_{-55}$   &   $39^{+42}_{-25}$ & $0.6^{+4.7}_{-0.5}$  & $0.3^{+0.1}_{-0.1}$ & $6.9^{+0.3}_{-0.4}$ & & $60^{+12}_{-12}$  \\
A2 & $25^{+8}_{-9}$   & $0.6^{+4.0}_{-0.6}$  & $0.1^{+0.1}_{-0.1}$ & $6.7^{+0.2}_{-0.2}$ & & $66^{+38}_{-29}$  &    $39^{+26}_{-17}$ & $0.3^{+3.4}_{-0.3}$  & $0.2^{+0.1}_{-0.1}$ & $6.7^{+0.2}_{-0.3}$ & & $108^{+66}_{-51}$  &   $48^{+33}_{-26}$ & $0.3^{+2.1}_{-0.2}$  & $0.2^{+0.1}_{-0.1}$ & $6.7^{+0.2}_{-0.3}$ & & $65^{+13}_{-13}$  \\
B1 & $20^{+14}_{-8}$  & $0.4^{+14.3}_{-0.4}$ & $0.6^{+0.1}_{-0.2}$ & $7.0^{+0.3}_{-0.3}$ & & $44^{+49}_{-22}$  &    $25^{+31}_{-14}$ & $1.2^{+31.5}_{-1.2}$ & $0.6^{+0.1}_{-0.2}$ & $6.9^{+0.4}_{-0.3}$ & & $83^{+102}_{-53}$  &   $34^{+59}_{-23}$ & $0.8^{+34.0}_{-0.8}$ & $0.6^{+0.1}_{-0.3}$ & $6.9^{+0.4}_{-0.4}$ & & $115^{+23}_{-23}$ \\
B2 & $12^{+6}_{-4}$   & $0.2^{+2.2}_{-0.2}$  & $0.6^{+0.1}_{-0.1}$ & $6.9^{+0.3}_{-0.3}$ & & $23^{+20}_{-10}$  &     $12^{+11}_{-5}$ & $2.5^{+50.9}_{-2.4}$ & $0.5^{+0.1}_{-0.3}$ & $6.7^{+0.3}_{-0.3}$ & & $36^{+42}_{-19}$   &    $14^{+17}_{-7}$ & $3.8^{+52.4}_{-3.6}$ & $0.5^{+0.1}_{-0.2}$ & $6.7^{+0.3}_{-0.3}$ & & $113^{+23}_{-23}$ \\
C1 & $13^{+14}_{-8}$  & $0.2^{+3.7}_{-0.2}$  & $0.4^{+0.2}_{-0.1}$ & $6.4^{+0.4}_{-0.3}$ & & $48^{+58}_{-30}$  &    $26^{+39}_{-16}$ & $0.4^{+5.4}_{-0.4}$  & $0.3^{+0.1}_{-0.2}$ & $6.5^{+0.4}_{-0.3}$ & & $59^{+82}_{-39}$   &   $23^{+42}_{-16}$ & $0.6^{+8.9}_{-0.6}$  & $0.4^{+0.1}_{-0.2}$ & $6.3^{+0.4}_{-0.3}$ & & $156^{+31}_{-31}$ \\
C2 & $21^{+28}_{-13}$ & $0.3^{+7.1}_{-0.3}$  & $0.4^{+0.3}_{-0.2}$ & $6.7^{+0.4}_{-0.4}$ & & $69^{+79}_{-48}$  &    $41^{+56}_{-29}$ & $0.5^{+9.9}_{-0.4}$  & $0.3^{+0.2}_{-0.2}$ & $6.7^{+0.4}_{-0.4}$ & & $77^{+97}_{-49}$   &   $32^{+57}_{-22}$ & $1.2^{+27.3}_{-1.1}$ & $0.3^{+0.2}_{-0.2}$ & $6.5^{+0.4}_{-0.4}$ & & $166^{+33}_{-33}$ \\
D1 & $30^{+62}_{-16}$ & $0.6^{+17.6}_{-0.6}$ & $0.5^{+0.2}_{-0.3}$ & $6.7^{+0.4}_{-0.5}$ & & $92^{+87}_{-54}$  &    $59^{+69}_{-36}$ & $0.7^{+19.8}_{-0.7}$ & $0.5^{+0.3}_{-0.3}$ & $6.6^{+0.4}_{-0.4}$ & & $117^{+117}_{-76}$ &   $54^{+80}_{-38}$ & $0.4^{+10.5}_{-0.3}$ & $0.6^{+0.2}_{-0.3}$ & $6.6^{+0.4}_{-0.5}$ & & $208^{+42}_{-42}$ \\
D2 & $50^{+68}_{-35}$ & $0.2^{+2.3}_{-0.2}$  & $0.2^{+0.2}_{-0.1}$ & $6.6^{+0.5}_{-0.6}$ & & $84^{+84}_{-54}$  &    $57^{+66}_{-38}$ & $0.2^{+5.6}_{-0.2}$  & $0.2^{+0.2}_{-0.2}$ & $6.5^{+0.5}_{-0.6}$ & & $111^{+132}_{-74}$ &   $54^{+87}_{-40}$ & $0.3^{+5.8}_{-0.2}$  & $0.3^{+0.3}_{-0.2}$ & $6.4^{+0.5}_{-0.5}$ & & $166^{+33}_{-33}$ \\
E2 & $11^{+10}_{-8}$  & $0.3^{+3.4}_{-0.2}$  & $0.4^{+0.3}_{-0.2}$ & $5.5^{+0.3}_{-0.4}$ & & $29^{+40}_{-18}$  &    $16^{+24}_{-10}$ & $0.8^{+24.2}_{-0.8}$ & $0.3^{+0.2}_{-0.2}$ & $5.4^{+0.5}_{-0.4}$ & & $54^{+68}_{-40}$   &   $21^{+34}_{-16}$ & $1.6^{+32.9}_{-1.5}$ & $0.3^{+0.2}_{-0.2}$ & $5.4^{+0.4}_{-0.4}$ & & $404^{+81}_{-81}$ \\
F1 & $38^{+78}_{-26}$ & $0.4^{+12.2}_{-0.4}$ & $0.5^{+0.4}_{-0.3}$ & $5.7^{+0.4}_{-0.5}$ & & $91^{+91}_{-61}$  &    $58^{+71}_{-42}$ & $0.7^{+11.1}_{-0.7}$ & $0.6^{+0.4}_{-0.4}$ & $5.6^{+0.4}_{-0.5}$ & & $107^{+141}_{-71}$ &   $50^{+88}_{-35}$ & $0.7^{+34.7}_{-0.6}$ & $0.5^{+0.4}_{-0.3}$ & $5.5^{+0.5}_{-0.5}$ & & $995^{+199}_{-199}$\\
F2 & $44^{+20}_{-15}$ & $0.2^{+1.5}_{-0.1}$  & $0.1^{+0.2}_{-0.1}$ & $5.4^{+0.2}_{-0.2}$ & & $118^{+59}_{-57}$ &    $78^{+42}_{-42}$ & $0.2^{+2.4}_{-0.1}$  & $0.2^{+0.3}_{-0.2}$ & $5.4^{+0.2}_{-0.2}$ & & $164^{+94}_{-89}$  &   $86^{+54}_{-54}$ & $0.5^{+6.4}_{-0.5}$  & $0.2^{+0.3}_{-0.2}$ & $5.4^{+0.2}_{-0.3}$ & & $995^{+199}_{-199}$\\
G1 & $22^{+44}_{-14}$ & $0.4^{+5.8}_{-0.4}$  & $0.3^{+0.3}_{-0.2}$ & $6.6^{+0.5}_{-0.5}$ & & $62^{+84}_{-42}$  &    $39^{+66}_{-29}$ & $0.4^{+11.2}_{-0.3}$ & $0.3^{+0.3}_{-0.2}$ & $6.6^{+0.5}_{-0.5}$ & & $74^{+108}_{-52}$  &   $32^{+67}_{-24}$ & $0.5^{+10.0}_{-0.4}$ & $0.3^{+0.3}_{-0.2}$ & $6.5^{+0.6}_{-0.5}$ & & $22^{+4}_{-4}$    \\
H2 & $67^{+87}_{-49}$ & $0.3^{+3.9}_{-0.3}$  & $0.3^{+0.4}_{-0.2}$ & $7.1^{+0.5}_{-0.6}$ & & $92^{+94}_{-68}$  &    $61^{+74}_{-48}$ & $0.5^{+7.2}_{-0.4}$  & $0.3^{+0.3}_{-0.2}$ & $6.8^{+0.6}_{-0.6}$ & & $124^{+132}_{-81}$ &   $61^{+95}_{-45}$ & $0.7^{+20.4}_{-0.6}$ & $0.3^{+0.3}_{-0.2}$ & $6.8^{+0.6}_{-0.6}$ & & $22^{+4}_{-4}$    \\
I1 & $62^{+73}_{-43}$ & $0.6^{+21.6}_{-0.6}$ & $0.5^{+0.5}_{-0.3}$ & $7.2^{+0.3}_{-0.4}$ & & $107^{+104}_{-67}$ &   $75^{+80}_{-51}$ & $0.6^{+22.3}_{-0.5}$ & $0.8^{+0.4}_{-0.5}$ & $7.2^{+0.3}_{-0.4}$ & & $136^{+114}_{-90}$ &   $66^{+82}_{-48}$ & $0.9^{+22.4}_{-0.9}$ & $0.9^{+0.3}_{-0.5}$ & $7.1^{+0.4}_{-0.4}$ & & $22^{+4}_{-4}$    \\
I2 & $95^{+122}_{-62}$& $0.5^{+6.3}_{-0.5}$  & $0.5^{+0.4}_{-0.3}$ & $7.3^{+0.5}_{-0.6}$ & & $139^{+138}_{-93}$ & $100^{+117}_{-73}$ & $0.7^{+27.4}_{-0.6}$ & $0.6^{+0.5}_{-0.4}$ & $7.2^{+0.6}_{-0.6}$ & & $181^{+178}_{-119}$&  $94^{+138}_{-70}$ & $1.0^{+17.6}_{-1.0}$ & $0.6^{+0.5}_{-0.4}$ & $7.1^{+0.6}_{-0.6}$ & & $22^{+4}_{-4}$    \\
J1 & $47^{+102}_{-27}$& $0.5^{+18.8}_{-0.4}$ & $0.7^{+0.5}_{-0.5}$ & $7.5^{+0.3}_{-0.3}$ & & $104^{+111}_{-66}$ &   $67^{+88}_{-45}$ & $0.9^{+16.0}_{-0.8}$ & $0.9^{+0.4}_{-0.5}$ & $7.4^{+0.3}_{-0.3}$ & & $139^{+145}_{-87}$ &   $69^{+95}_{-49}$ & $0.4^{+15.6}_{-0.4}$ & $1.1^{+0.3}_{-0.5}$ & $7.5^{+0.3}_{-0.3}$ & & $22^{+4}_{-4}$    \\
J2 & $47^{+82}_{-29}$  & $0.3^{+7.5}_{-0.3}$  & $0.2^{+0.3}_{-0.2}$ &$7.0^{+0.4}_{-0.4}$  & & $107^{+92}_{-60}$ &   $68^{+77}_{-40}$ & $0.2^{+3.4}_{-0.2}$  & $0.3^{+0.3}_{-0.2}$ & $7.0^{+0.4}_{-0.4}$ & & $142^{+114}_{-92}$ &   $66^{+78}_{-47}$ & $0.4^{+10.0}_{-0.3}$ & $0.4^{+0.3}_{-0.2}$ & $6.9^{+0.4}_{-0.4}$ & & $22^{+4}_{-4}$    \\

\bottomrule
\addlinespace[3pt]
\multicolumn{19}{p{\textwidth}}{\textbf{Note.} $t_\star$ and $t_{\rm{MW}}$ are in Myr, $Z$ is in units of $\% Z_\odot$, and $A_V$ is expressed in mag. Errors correspond to the 16th and 84th percentiles of the posterior distribution. $\mu$ represents the lensing magnification factor and results from averaging the two lensing models' estimates from \citetalias{messa_jwst_2026}, as described in the text.}
\end{tabular}
\end{table*}

\subsection{Comparison with state of the art}\label{sec:App_comparison}

\begin{figure*}
    \centering
    \includegraphics[width=0.98\linewidth]{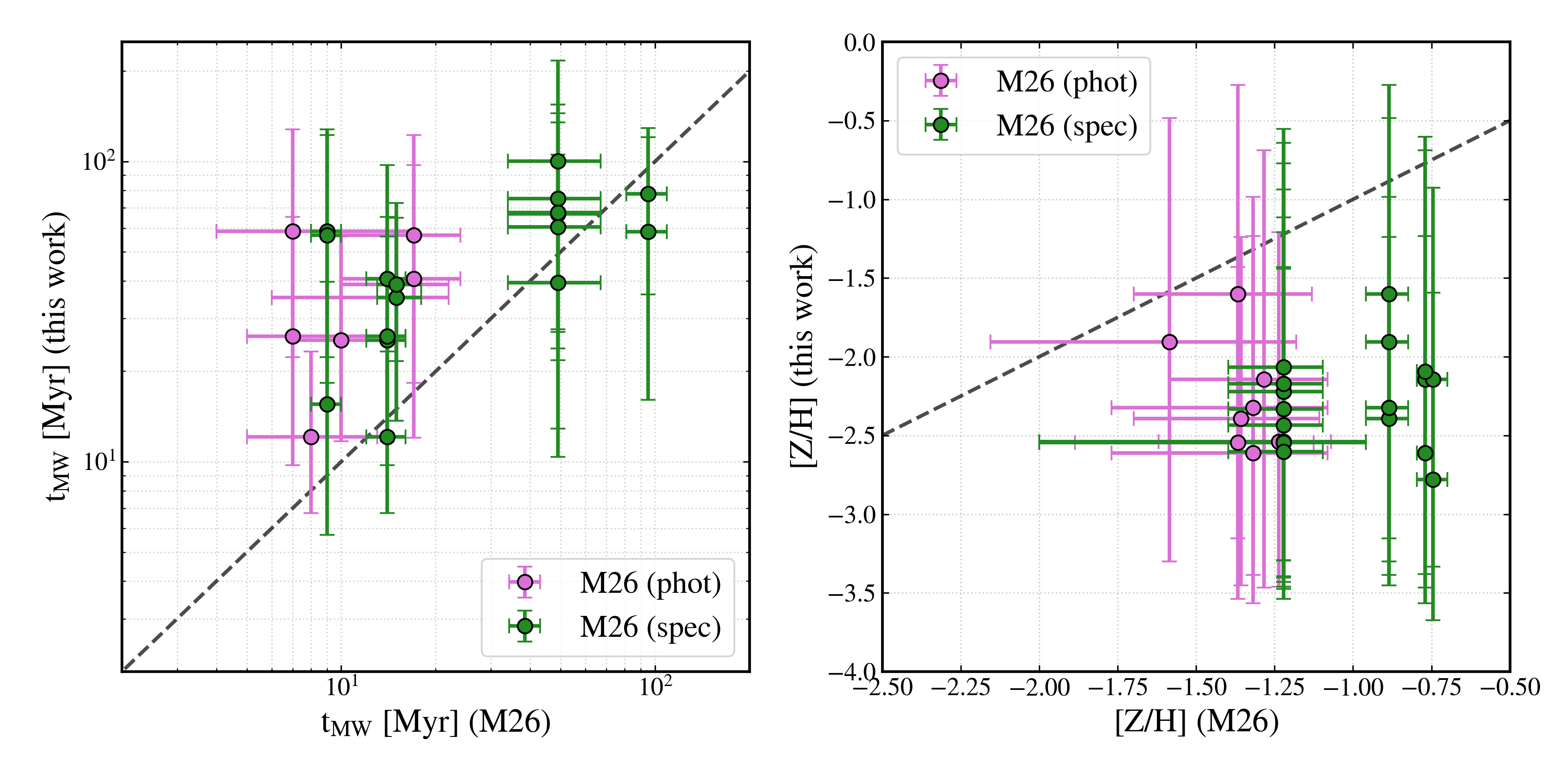}
    \caption{Mass-weighted ages (top) and metallicity (bottom) comparison with \citetalias{messa_jwst_2026} results. On the x-axis, we report the results from their photometric analysis (in pink), and the ones from the spectroscopic fit (in green). While the results from photometry are limited to the central sources (A1-E1, A2-D2), the spectroscopic ones include also the tail clusters and the innermost region (CC region in \citetalias{messa_jwst_2026}). Values on the y-axis refer to the results obtained in this work with the Exponential SFH configuration.}
    \label{fig:comparison}
\end{figure*}
We compare our results with the initial photometric study by \citetalias{adamo_bound_2024}, which focused on the central star clusters (A–E), and the recent spectroscopic analysis by \citetalias{messa_jwst_2026}. The latter was performed using masks covering the central sources (split into A1,2 and BCDE1,2), a central region (CC) aligning with our sources F1–F2, and a tail mask encompassing the external region (our G–J). 

While our analysis uses \texttt{BAGPIPES} along with the \citetalias{Bruzual2003} templates, \citetalias{adamo_bound_2024} explored both \texttt{BAGPIPES} and \texttt{Prospector} \citep{johnson_stellar_2021}, implementing separate analyses with BPASS v2.2.1 \citep{stanway_re-evaluating_2018} to account for stellar binaries. Similarly, \citetalias{messa_jwst_2026} opted for BPASS models as their primary reference. 

In Fig. \ref{fig:comparison}, we show the comparison of mass-weighted ages (upper panel) and metallicities (lower panel) obtained in this work with respect to the results obtained in \citetalias{messa_jwst_2026}, both from the photometric and spectroscopic analyses. Despite the differences in the SSP templates adopted, the resulting mass-weighted ages remain remarkably consistent within uncertainties. Our estimates are, on average, only $15$ Myr and $24$ Myr older than \citetalias{messa_jwst_2026} spectroscopic and photometric estimates, respectively. Metallicities converge instead towards systematically lower values, which nevertheless remain compatible within the large error bars. Specifically, we find average negative offsets of $-0.9$ dex and $-1.2$ dex relative to the photometric and spectroscopic measurements, respectively.

For the central star clusters A-E, our mass-weighted ages range from 11 to 50 Myr in the burst model and 12 to 59 Myr in the exponential and delayed configurations. This is in good agreement with the 9–36 Myr window reported by \citetalias{adamo_bound_2024}, and our younger sources also match the spectroscopic age constraints of 12–15 Myr derived from the corresponding masks in \citetalias{messa_jwst_2026}. 

Regarding the innermost region (sources F1–F2), which corresponds to the central CC mask in \citetalias{messa_jwst_2026}, we find that they are more evolved than the core clusters, with mass-weighted ages spanning from 38–45 Myr (burst) up to 50–86 Myr (exponential/delayed). This is consistent with the mass-weighted age of $95 \pm 14$ Myr derived spectroscopically by \citetalias{messa_jwst_2026}.

Finally, the external tail region (sources G–J) maps directly onto the tail mask analysed by \citetalias{messa_jwst_2026} and hosts the oldest populations in the arc, with our mass-weighted ages falling in the range 22-100 Myr. This significant age gradient supports the scenario outlined by \citetalias{messa_jwst_2026}, who characterized the tail as having a more prolonged SFH and an older overall mass-weighted age ($\sim$43–68 Myr) compared to the central, bursty clusters. 

Our derived mean stellar metallicities, $Z = 0.4 - 0.9 \% \: Z_\odot$, and their low dust content, $A_V\sim 0.3-0.4$ mag, agree well with the photometric estimates from \citetalias{adamo_bound_2024} under their single-burst assumption, where they reported $Z < 1\%\,Z_\odot$ and $A_V < 0.3$ mag. However, when more complex SFHs are assumed by \citetalias{adamo_bound_2024}, as well as in the spectroscopic results from \citetalias{messa_jwst_2026}, the chemical enrichment is found to be higher, reaching levels of 10\%–20\% $Z_\odot$, while still predicting low dust extinction. To test this further, we perform an alternative fit restricting the stellar metallicity to a range of 3\%–30\% $Z_\odot$, closer to the values suggested by spectroscopy. This leads to a shift of approximately 20 Myr in the age peak of our older components; although this variation represents a large percentage change, it still falls comfortably within the large statistical errors of our primary fit, indicating that the choice of metallicity prior does not critically affect our conclusions.

\end{appendix}

\end{document}